\documentclass[lettersize,journal]{IEEEtran}
\usepackage{amsmath,amsfonts}
\usepackage{amsthm}
\usepackage{amssymb}
\usepackage{array}
\usepackage[caption=false,font=normalsize,labelfont=sf,textfont=sf]{subfig}
\usepackage{textcomp}
\usepackage{stfloats}
\usepackage{url}
\usepackage{verbatim}
\usepackage{graphicx}
\usepackage{cite}
\usepackage{amsfonts} 
\usepackage[short]{optidef}
\allowdisplaybreaks[4]
\usepackage[subtle]{savetrees}
\usepackage{bbm}
\usepackage{pdfpages}  
\usepackage{epstopdf}

\usepackage[normalem]{ulem}
\newcommand{\breal}[1]{\overline{\underline{#1}}}
\newcommand{\areal}[1]{\underline{#1}}
\newcommand{\bimag}[1]{\overline{\dotuline{#1}}}
\newcommand{\aimag}[1]{{\dotuline{#1}}}
\newcommand{\asub}[1]{\substack{#1\\[1.0ex]~}}

\usepackage{mathtools}
\usepackage{setspace}
\allowdisplaybreaks
 
\usepackage[ruled,vlined]{algorithm2e} 

\def\nby{{\mathbf{y}}}

\def\nb0{{\mathbf{0}}}
\def\nb1{{\mathbf{1}}}

\def\nbbC{{\mathbb{C}}}

\def\nbbP{{\mathbb{P}}}

\def\nbbR{{\mathbb{R}}}

\def\nbbZ{{\mathbb{Z}}}

\newtheorem{lem}{Lemma}

\newtheorem{nrem}{Remark}

\newtheorem{eg}{Example}

\usepackage{enumitem}
\renewcommand{\labelenumii}{\arabic{enumi}.\arabic{enumii}}
\renewcommand{\labelenumiii}{\arabic{enumi}.\arabic{enumii}.\arabic{enumiii}}
\renewcommand{\labelenumiv}{\arabic{enumi}.\arabic{enumii}.\arabic{enumiii}.\arabic{enumiv}}
\begin{document}
\title{\vspace{-0.3em}Input Distribution Optimization in OFDM Dual-Function Radar-Communication Systems}
\author{Yumeng Zhang, Sundar Aditya,~\IEEEmembership{Member,~IEEE} and Bruno Clerckx,~\IEEEmembership{Fellow,~IEEE} 
\thanks{This work has been partially supported by UKRI grant EP/X040569/1, EP/Y037197/1, EP/X04047X/1, EP/Y037243/1.}
\thanks{The authors are with the Department of Electrical and Electronic Engineering, Imperial College London, London SW7 2AZ, U.K. (e-mail:
\{yumeng.zhang19, s.aditya, b.clerckx\}@imperial.ac.uk).}
\vspace{-0.3em}}
\maketitle
\begin{abstract}
Orthogonal frequency division multiplexing (OFDM) has been widely adopted in dual-function radar-communication (DFRC) systems. However, with random communication symbols (CS)  {embedded in the DFRC waveform}, the transmit signal has a random ambiguity function that affects the radar's delay-Doppler estimation performance,  {which has not been well explored}.  {This paper addresses this gap by first characterizing the  outlier probability (OP) -- the probability of incorrectly estimating a target's (on-grid) delay-Doppler bin -- in OFDM DFRC \textcolor{blue}{for any given CS realization}.  This subsequently motivates the OFDM DFRC waveform design problem of minimizing the OP w.r.t the CS probability distribution (i.e., the \textit{input distribution}). } Conditioned on the CSs, the OP only depends on the CS magnitudes. Hence, we consider the following two schemes for the above optimization: CSs with (1) constant magnitude input distribution (phase shift keying), and (2) variable magnitude input distribution (Gaussian). For (1),  {minimizing the OP} reduces to the familiar power allocation design across OFDM's subcarriers and symbols, with uniform power allocation across  {OFDM} subcarriers and a \textit{windowed} power allocation across  {OFDM} symbols being near-optimal. For (2), the mean and variance of the Gaussian distribution at each subcarrier is optimized, with an additional communication constraint to avoid the zero-variance solution where no CSs are carried. We observe that subcarriers with strong communication channels feature   {a large} variance (favour communications) while the others are characterized by a  {large} mean (favour radar). However, the overall power allocation (i.e., the sum of  {the squared} mean and variance) across the OFDM subcarriers and symbols is similar to (1).  {Simulations for (2)  {show that while random } CS magnitudes benefit communications, they degrade radar performance, but this can be mitigated using our optimized input distribution.}

\end{abstract}

\begin{IEEEkeywords}
OFDM signal design, dual-functional radar-communication, random communications symbols, input distribution
\end{IEEEkeywords}

\section{Introduction}
{Dual-function} radar-communication (DFRC) has attracted considerable interest as a solution to the problem of spectrum shortage arising from the increasing use of wireless technologies \cite{paul2016survey,8999605}. In DFRC, a common signal is used for radar and communications simultaneously \cite{9737357,9705498}. Hence, of vital significance in DFRC systems is the choice and design of a suitable signal that performs both communications and radar satisfyingly \cite{hwang2008ofdm,sturm2011waveform}. 

The orthogonal frequency division multiplexing (OFDM) signal is a strong candidate for DFRC systems given its benefits for both communications and radar. For the former, OFDM is well-known to provide robustness against channel fading and combat inter-symbol interference to realize higher data rates \cite{hwang2008ofdm}. For the latter, a highly efficient 2D-FFT (2-dimensional fast Fourier transform) based delay-Doppler bin estimator exists and has been shown to achieve very promising estimation accuracy \cite{sturm2011waveform,8805161,berger2010signal}. 

The performance of OFDM for communications and radar in DFRC can be further enhanced through waveform optimization (e.g., optimizing the power allocation across subcarriers) \cite{liyanaarachchi2021optimized,7970102,sen2009adaptive}. The communications/radar metrics adopted for such optimization should be both intuitive and tractable. A widely accepted metric for communications is the achievable rate.  {For radar}, the most recognized metrics include: (a) the detection probability (DP)/false alarm probability (FAP) that describes the probability of correctly/falsely detecting the existence of
a target in a certain delay/Doppler/angle bin \cite{1597550, moulin2022joint, chen2018waveform}, and (b) the mean square error (MSE) of parameter estimation \cite{athley2005threshold}. However, these  {radar} metrics usually result in high complexity when adopted for waveform optimization. As a substitute, indirect metrics based on desirable properties that {good} sensing signals often share -- such as the ambiguity function (AF) \cite{sen2009adaptive, lellouch2016design,barneto2021beamformer,9473742,9771644} -- have also been used.  {Other indirect metrics that have been adopted for OFDM  {waveform optimization} include signal-to-noise-ratio (SNR) or the signal-to-interference-and-noise-ratio (SINR) as a substitute for DP/FAP\cite{shao2022target,9906898}; radar mutual information (RMI) from an information theory perspective \cite{259642,7970102}; or the Cramer-Rao bound (CRB) as a lower bound of MSE\cite{dogandzic2001cramer,9705498,liyanaarachchi2021optimized}.}

 {When using any radar metric for OFDM DFRC waveform optimization, it is important to account for the inherent randomness in the waveform due to the embedded} information-bearing communications symbols (CS) -- e.g., QAM. {However,  previous literature on waveform optimization using the above metrics does not fully capture the impact of the CS probability distribution -- henceforth termed the \textit{input distribution} -- on radar performance. }\textcolor{blue}{For instance, maximizing an OFDM waveform's average SNR -- an indirect metric for optimizing DP/FAP -- only optimizes power allocation across subcarriers. This optimization often assumes that all power (equal to the sum of \textit{the squared mean} and the variance) is allocated to the variance of subcarriers' CSs for communications, while assuming zero means \cite{9906898,barneto2021beamformer}.} \textcolor{blue}{This ignores the detrimental role that the  {variance of the input distribution} plays in radar performance,  {as for radar, it is sufficient to have a deterministic signal, represented by the mean of the input distribution with zero variance}. Hence, in DFRC, it is inadequate to optimize solely the power allocation, which is captured by the second moment of the input distribution. In contrast, it is also significant to determine how the power is allocated between the mean and the variance to favour both radar and communications.}  {Similar issues exist in the literature for other sensing metrics}, such as the RMI  where the average RMI of random CSs is substituted by the RMI of the average CSs \cite{7970102,bica2015opportunistic}.


The effect of input distribution on radar is worth considering, since there is ample literature on the detrimental effects of
different sources of randomness in radar systems (e.g., 
the fluctuations of the target's radar cross-section, and signal-dependent clutter \cite{1597550,levanon2004radar,karimi2019adaptive}), and how performance gains can be achieved by accounting for such randomness through detector/waveform design \cite{1597550,karimi2019adaptive}. Regarding the input distribution, \cite{hu2014radar} has demonstrated that the phase randomness of CSs has only a trivial effect on the OFDM waveform's AF, which makes a compelling case for using constant magnitude phase-coded CSs (e.g., PSK symbols) in OFDM waveforms for DFRC to achieve good radar performance \cite{8378693, skolnik2008radar, tian2017radar, aditya2022sensing}. However, to support higher communication rates, input distributions with varying magnitudes (e.g., 256-QAM) may need to be embedded into an OFDM waveform, and the impact of such distributions on radar performance has not been explored.  

 \textcolor{blue}{Very recently, \cite{xiong2023fundamental,liu2023deterministic} shed light on the radar-communication (R-C) trade-off in DFRC, introduced by the randomness of the input distribution. The authors characterized the R-C region (the best achievable radar performance given communication constraints), and examined the optimal input distributions at the two corner points on the R-C region (i.e., radar-optimal and communication-optimal respectively).  The paper reveals the fundamental R-C trade-off that the radar-optimal input distribution features CSs with deterministic sample covariance (constant-magnitudes) while the communication-optimal input distribution features CSs with zero-mean Gaussian (random-magnitudes).  However, for the practically relevant problem in DFRC systems of on-grid delay-Doppler bin estimation, the metrics used in  \cite{xiong2023fundamental} (radar MI) and \cite{liu2023deterministic} (CRB) may not be the most suitable. For instance, the radar MI metric in  \cite{xiong2023fundamental} is best suited for extended targets and lacks efficient solutions for arbitrary communication constraints. Similarly, the CRB in \cite{liu2023deterministic} is best suited for continuous-valued parameters and introduces significantly higher complexity as the number of estimation parameters increases. Therefore, for the on-grid delay-Doppler bin estimation in DFRC, a more appropriate and tractable metric is needed for waveform design, as the parameter space is discrete and two-dimensional.} 

In this paper, we  {consider an OFDM DFRC waveform design via input distribution optimization with the radar task being on-grid delay-Doppler bin estimation, which is of considerable practical relevance.}  Our contributions are summarized as follows:

  {(1)  {We start from why the outlier probability\footnote{ {OP is defined as the probability that a wrong delay-Doppler bin is estimated. For the special case where it is known apriori that a target is present in one of the delay-Doppler bins, the OP coincides with the FAP, i.e., the 'false alarm' refers to deciding the target to any of the wrong bins after detection.}}  (OP) is a suitable metric for characterizing the on-grid delay-Doppler bin estimation performance of the maximum likelihood (ML) estimator.  We then derive an upper bound for the OP (UBOP) that is a good approximation of the OP across the whole SNR region. We show that the UBOP coincides with the signal's AF properties and is only a function of the CS magnitudes. The UBOP is adopted as the radar metric throughout the paper.} }

     {(2) We minimize the UBOP w.r.t the input distributions of CSs with constant magnitudes  (e.g. PSK), to provide intuition for radar-favoured signals. In this scheme, information is only conveyed through the CS phases, and the radar performance of the optimized solution matches that of a  {deterministic constant envelope} signal in a pure radar system. \textcolor{blue}{We propose efficient closed-form solutions for the optimization problem,} {whose optimal input distribution features uniform power allocation across the OFDM subcarriers (to favour delay estimation) and a windowed power allocation across the OFDM symbols (to favour Doppler estimation).}  { This is also the first paper to   {design signals} across both the time and frequency domains because of a joint consideration of delay-Doppler estimation.} This scheme is hence used as a benchmark for the  {case of} CSs with variable magnitude input distributions.}  

     {(3)  {We} consider the DFRC performance for input distributions of CSs with variable magnitudes, by minimizing the average UBOP (aUBOP) w.r.t the \textit{non-zero} mean Gaussian input distribution, \textcolor{blue}{given arbitrary communication rate constraints.} Deviating from the intuitive choice of a \emph{zero-mean} Gaussian input distribution for communications, this is the first paper to optimize a \emph{non-zero} Gaussian input distribution for DFRC systems. A non-zero mean Gaussian gives rise to a trade-off in allocating the available power budget between the symbol mean and the symbol variance,  with radar/communications favouring a less/more random input distribution through greater power allocation to the symbol mean/variance. More specifically, the simulations reveal that for subcarriers having high channel gain for communications, power is concentrated on the symbol variance in order to satisfy the communication constraints efficiently. On the other hand, for subcarriers with weak channel gains for communications, power is concentrated on the symbol means to favour radar performance. The power allocation (sum over the squared symbol means and variance) across the OFDM subcarriers, however, is similar to (2).  \textcolor{blue}{A simplified algorithm is also proposed for this scheme to handle large dimensional data, by decoupling the OFDM symbol optimization across the time domain and the frequency domain.} }


\textit{Organization:} 
Section~\ref{sec:model} introduces the signal and system model of a point-target mono-static SISO DFRC scenario  {using OFDM waveform}. Then, Section~\ref{sec:radar_rx} develops the ML estimator for joint delay-Doppler bin estimation, derives the OP conditioned on CS  magnitudes and its tractable upper bound (UBOP). Section~\ref{sec_BPSK} considers  {the problem of minimizing the UBOP for PSK input distribution to shed insight into the patterns of power allocation across OFDM subcarriers and OFDM symbols that yield good radar performance.} Section~\ref{sector_Gaussian}  {extends the scope of input distributions to those symbols with variable magnitudes, by minimizing the aUBOP subject to a communication achievable rate constraint for a non-zero mean Gaussian input distribution.}  Section~\ref{sec:simulations} evaluates the OP performance of  {the optimized input distributions from Sections \ref{sec_BPSK} and \ref{sector_Gaussian}} and finally, Section~\ref{sec:concl} concludes the paper. 

\textit{Notation:}  
Throughout the paper, matrices and vectors are respectively denoted in bold upper case and bold lower case. $\nbbR$, $\nbbC$ and $\nbbZ$ denotes the set of real numbers, complex numbers and integers, respectively. For $x \in \nbbC$, $\mathfrak{R}\{x\}$, $\mathfrak{I}\{x\}$, $\measuredangle x$ and $|{x}|$ represents the real part, the imaginary part, the phase and the magnitude of ${x}$, respectively. For a vector (matrix) $\mathbf{x}$ ($\mathbf{X}$), $\|\mathbf{x}\|$ ($\|\mathbf{X}\|$) represents its $l_2$ (Frobenius) norm, $x_n$ refers to the $n^{\mathrm{th}}$ entry of vector $\mathbf{x}$, and $\mathrm{diag}\{\mathbf{x}\}$($\mathrm{diag}\{\mathbf{X}\}$) represents  {the diagonal matrix (vector) formed by $\mathbf{x}$ (matrix $\mathbf{X}$).} $|\mathbf{x}|$  {denotes the} element-wise magnitude of $\mathbf{x}$, i.e., $|\mathbf{x}|_k= |{x}_k|$.  $\mathbf{I}_{N_t}$ denotes $N_t\times N_t$ an identity matrix and $\mathbf{1}_K$ is the all-one column vector of length $K$ (the size of the vector may not be explicitly specified where it is obvious). $(\cdot)^H$, $(\cdot)^*$ and $(\cdot)^T$ represent the Hermitian, the conjugate and the transpose operations respectively. $\odot$ represents the element-wise (Hadamard) product, $\oslash$ represents the element-wise (Hadamard) division, $\otimes$ represents the Kronecker product. Given a set $\kappa$, $|\kappa|$ refers to the cardinality of $\kappa$. For $x \in \nbbR$, $(x)^+ \triangleq \max( x, 0)$, and $\mathbbm{1}_{(\cdot)}$ is the indicator function. For a random variable $X$, $\mathbb{E}_X\left\lbrace f(X)\right\rbrace$ represents the expected value of the function $f(X)$ over $X$. $\nbbP(\cdot)$ denotes probability. $Q(\cdot)$ is the Marcum-Q function, and $I_0(\cdot)$ is the modified Bessel function of the first kind with order 0. $\mathcal{N}(\mu,~\sigma^2)$ ($\mathcal{CN} (\mu,~\sigma^2)$) refers to the normal (complex normal) distribution with mean $\mu$ and variance $\sigma^2$ (for the complex normal, the real and imaginary parts are independent with variance $\sigma^2/2$). 

\section{Signal Model}
\label{sec:model}
This section models a DFRC system where an OFDM transmit signal is used to simultaneously communicate with a user and sense a point target. We consider an OFDM block containing $M$ OFDM symbols. Each OFDM symbol has $K$ subcarriers, with an additional $K_G$ sub-pulses of cyclic prefix (CP). We assume that a target's delay and Doppler remain unchanged over one block.

Let $B$ denote the signal bandwidth, corresponding to a delay resolution of $\Delta \tau=1/B$. We assume that the target round-trip delay is shorter than the CP duration (the maximal delay that can be estimated is $\tau_{\max}={K_G/B}$). Let $T_{\mathrm{OFDM}}=(K+K_G)/B \triangleq K_T/B$ denote the OFDM symbol duration. Hence, the radar coherent {processing} interval (CPI) equals $M T_{\mathrm{OFDM}}$, corresponding to  the Doppler resolution of $\Delta f_{D}=1/(MT_{\mathrm{OFDM}})=B/(K_TM)$.




Consider the presence of a target at a distance $R$ and moving at velocity $v$ w.r.t the radar. For a monostatic configuration, $R$ and $v$ are captured on the echo signal by its time delay $\tau = 2R/c$ and Doppler frequency $f_{\mathrm{D}} = f_c (2v/c)$,  {with $c$ being the speed of light}. $\tau$ and $f_{\mathrm{D}}$  can be represented as follows
\begin{align}
\label{eq_resolution}
\tau &= (n_0+n_{\epsilon}) \Delta \tau, \hspace{3mm}n_0 \in \{0,~1,~\cdots,~ K_G-1\}, \\
\label{eq_resolution1}
 f_{\mathrm{D}} &=(v_0+v_{\epsilon}) \Delta f_{\mathrm{D}}, \hspace{3mm} v_0\in  \{-{M}/{2},~\cdots,~0,~\cdots,~{(M-1)}/{2} \}, 
\end{align}
{where} $ n_{\epsilon} \in (0,1)$ and $~v_{\epsilon} \in\left(-1/2,~1/2\right)$ are the fractional  {(off-grid)} delay and Doppler parameters.  {$(n_0,~v_0)$ denotes the target's  {\emph{on-grid} } delay-Doppler bin,  which is the parameter of interest in this paper.} 

Let $X[k,~m]$ denote the CS at the $k^{\mathrm{th}}~(k = 0,~ \cdots,~K-1)$ subcarrier of the $m^{\mathrm{th}}~(m = 0,~\cdots,~M-1)$ symbol. After a $K$-point IDFT, let $x[n,~m]~(n = 0,~ \cdots,~K-1)$ denote the resulting time domain signal. After adding the CP, we obtain the following signals
\begin{equation}
\label{eq_CP_discrete}
x_{\mathrm{CP}}[n,~m]=\begin{cases} x[K-K_G+n,~m],& n \in \{0, ~\cdots, ~K_G-1\}, \\ x[n-K_G,~m],&  n \in \{K_G, ~\cdots, ~K_T-1\},\end{cases}
\end{equation} 
 {The (continuous-time) signal transmitted over the air is given by \cite{lin2003ofdm}}
\begin{align} 
    \label{eq_CP_x_t}
    x(t)=\sum_{n,m}x_{\mathrm{CP}}[n,~m]\mathrm{rect}\left( t-(n+K_T m)/B\right).
\end{align}
The radar return at the transceiver (after ADC) is given by 
\begin{subequations}\begin{align}
&  y_{\mathrm{CP}}[n,~m]=ax(t-\tau)e^{j2\pi f_{\mathrm{D}}t}|_{t=(n+K_T m)/B}\\\label{eq_y_cp}
=& ax_{\mathrm{CP}}[n-n_0,~m]e^{j2\pi\frac{f_{\mathrm{D}}(n +mK_T)}{B}}  + w[n,~m]  \\ \label{eq_y_cp1}
=& a x_{\mathrm{CP}}[n-n_0,~m]e^{j2\pi \frac{f_{\mathrm{D}} n}{B}} e^{j2\pi \frac{(v_0 + v_{\epsilon})m}{M}} + w[n,~m] \\\label{eq_y_cp2}
\approx& a x_{\mathrm{CP}}[n-n_0,~m]e^{j2\pi  \frac{(v_0 + v_{\epsilon})m}{M}} + w[n,~m],
\end{align}
\end{subequations}
where (i) $a \in \nbbC$ is the scattering coefficient of the target, assumed to be unchanged throughout the OFDM block,  {(ii) in \eqref{eq_y_cp}, the received samples after ADC lack information on the fractional $ n_{\epsilon} $ because of the rectangular basis in \eqref{eq_CP_x_t}. Hence, $n_{\epsilon}$ is often disregarded in OFDM radar and the target delay is assumed to be on-grid \cite{mercier2020comparison}},  (iii) $e^{j2\pi f_{\mathrm{D}}n/B}  \approx 1$, since typically $f_{\mathrm{D}} \ll B$, and (iv) $w[n,~m]\sim\mathcal{CN}(0,~\sigma_{\mathrm{N}}^2)$ is the AWGN.  After removing the CP, we obtain the following receive signals
\begin{align}
\label{eq_rm_cp}
y[n,~m] &\approx a x[n-n_0,~m]e^ {j2\pi \frac{(v_0+v_{\epsilon})m}{M}}+w[n,~m],
\end{align}
where $x[n-n_0,~m]$ represents a circular shift of $n_0$ following \eqref{eq_CP_discrete}. After a $K$-point DFT, we get the frequency-domain signal as follows
\begin{subequations}
\begin{align}
\label{eq_receive_signal_DFT}
Y[k,~m] &=  \mu_{k,m}(\mathrm{\boldsymbol{\theta}}_0) + W[k,~m], \\
\mbox{with}~\label{eq_receive_signal_DFT2}
\mu_{k,m}(\mathrm{\boldsymbol{\theta}}_0) &:{=} a X[k,~m]e^{ {-j2\pi\frac{n_0k}{K}} } e^{ {j2\pi \frac{(v_0+v_{\epsilon})m} {M}} }, \\
\label{eq_receive_signal_DFT3}
W[k,~m] &\sim \mathcal{CN}(0,~\sigma_{\mathrm{N}}^2).
\end{align}
\end{subequations}
In (\ref{eq_receive_signal_DFT})-(\ref{eq_receive_signal_DFT3}), $\mathrm{\boldsymbol{\theta}}_0\triangleq [n_0, ~v_0, ~v_{\epsilon}, ~A_0, ~\phi_0]^T$ denotes  {the true values} of all of the unknown parameters in the system, with $A_0$ and $\phi_0$ denoting the amplitude and phase of $a$.  $n_0$ and $v_0$ are the parameters of interest, whereas the others are nuisance parameters.

\section{Characterizing the ML estimator's OP}
\label{sec:radar_rx}
This section begins by formulating the ML estimator for $(n_0, ~v_0)$ over an OFDM block. We then derive the OP of this estimator. The OP is then approximated by its upper bound to yield a tractable metric that can be optimized w.r.t the CS input distribution. 
 
\subsection{ML estimator}
\label{subsec:ML_estm}
From \eqref{eq_receive_signal_DFT}-\eqref{eq_receive_signal_DFT3}, the joint probability density function (PDF) of $Y[k,~m]$, conditioned on the unknown parameters $ {\boldsymbol{\theta}}\triangleq [ {n}, ~ {v}, ~ {v}'_{\epsilon}, ~{A}, ~{\phi}]^T$, is given  by
\begin{align}
\label{eq_pdf_Y_{m,k}}
f(\mathbf{y}| {\boldsymbol{\theta}})  
=  {{(\pi\sigma_{\mathrm{N}}^2)}^{-KM}} ~ e^ { - \sum_{m=0}^{M-1}\sum_{k=0}^{K-1} \big|Y[k,~m]-\mu_{k,m}( {\boldsymbol{\theta}}) \big|^2 \big/\sigma_{\mathrm{N}}^2},
\end{align}
where $\mathbf{y} \triangleq [\mathbf{y}^T_0, ~ \cdots, ~\mathbf{y}^T_{M-1}]^T$ with $\mathbf{y}_m \triangleq [Y[0,~m], ~\cdots,~ Y[K-1,~m]]^T$. The ML estimate of $ {\boldsymbol{\theta}}_0$, denoted by $\hat{\boldsymbol{\theta}}$, is given  by
\begin{subequations}\label{eq_ML0}
\begin{align}
\hat{\boldsymbol{\theta}} =& \arg \max_{ {\boldsymbol{\theta}}}~ \log f(\nby| {\boldsymbol{\theta}}) \\
\nonumber
=&\arg \max_{ {\boldsymbol{\theta}}}~ 2 {A} \mathfrak{R}\Big\{ e^{-j {\phi}}\sum_{m,k} Y[k,~m]X[k,~m]^*e^{j2\pi\frac{ {n} k}{K}}e^{-j2\pi \frac{( {v} +  {v}'_{\epsilon}) m} {M}}\Big\} \\\label{eq_pdf_Log2}
&-\sum_{m,k}\left[|Y[k,~m]|^2+| {A} X[k,~m]|^2 \right], 
\end{align}
\end{subequations}

\begin{nrem}
\label{rem:ve}
 {In \eqref{eq_ML0}, estimating the off-grid parameter $v_\epsilon$ (embedded in $Y[k,~m]$) typically involves exploiting sparsity (of targets), and is attractive because it provides "super-resolution". {In this paper however, we restrict our focus to on-grid parameter estimation (with its well-defined resolution limits \cite{mercier2020comparison,sturm2009novel}), as it is a relatively simpler starting point to investigate the relationship between the input distributions and sensing performance.} Hence, from the following ML estimator, we treat $v_{\epsilon}$ as a nuisance parameter affecting the estimation of on-grid parameters, $(n_0,~ v_0)$ (neglecting $v_{\epsilon}$), at the processor in \eqref{eq_ML0}. 
}
\end{nrem}

 {In \eqref{eq_pdf_Log2}, only the first term is a function of the target's delay-Doppler bin. Hence, the ML estimator of $(n_0,~v_0)$ is given by }(Assume the true delay-Doppler bin $(n_0,~v_0) = (0,~0)$ without loss of generality \cite{kay1993fundamentals})
\begin{subequations}
\begin{align}
\label{eq_ML_nv}
(\hat{n},~\hat{v})
=&\arg \underset{(n,~v)}{\max} ~ \Big|\sum_{m,~k} Y[k,~m]X[k,~m]^*e^{j2\pi\frac{nk}{K}}e^{-j2\pi \frac{v m} {M}} \Big| \\\label{eq_ML_reformulate1}
=&  \arg \underset{(n,~v)}{\max} ~ \Big|\sum_{m,~k} A p_{k,m}e^{j2\pi\frac{nk}{K}}e^{-j2\pi \frac{v m} {M}}e^{j2\pi \frac{v_{\epsilon} m} {M}}+w_\mathrm{o}[n,~v]\Big|\\
\label{eq_ML_re-formulate2}
=&\arg\underset{(n,~v)}{\max}\:\: |\gamma_{n,v,v_{\epsilon}}|, \\\nonumber
\mbox{where}& ~ \\\label{eq_ML_reformulate_define2}
p_{k,m} &= |X[k,~m]|^2 ,\\\label{eq_ML_reformulate_define3}
\gamma_{n,v,v_{\epsilon}} &= r_{n,v,v_{\epsilon}}+w_\mathrm{o}[n,~v] ,\\
\label{eq_ML_reformulate_define}
r_{n,v,v_{\epsilon}} &= \sum_{m,k} p_{k,m}e^{j2\pi\frac{nk}{K}}e^{-j2\pi \frac{v m} {M}}e^{j2\pi \frac{v_{{\epsilon}}m} {M}}, \\
\label{eq_ML_reformulate_define1}
w_{\mathrm{o}}[n,~v] &= {A}^{-1}\sum_{m,~k} W[k,~m]X[k,~m]^*e^{j2\pi\frac{nk}{K}}e^{-j2\pi \frac{v m} {M}},
\end{align}
\end{subequations}
for $n \in \{0,~\cdots, ~ K_G - 1\}, ~ v \in \{-M/2,~ \cdots, ~M/2-1 \} $. 

\begin{figure}[t]
\centering
\includegraphics[width=2.2in]{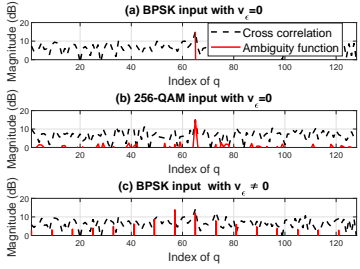}
\caption{ {The magnitudes of discrete AF  {($|r_{n,v,v_{\epsilon}}|$ in \eqref{eq_ML_reformulate_define})} and the corresponding cross correlation  {($|\gamma_{n,v,v_{\epsilon}}|$ in \eqref{eq_ML_reformulate_define3} averaging over random CSs for (b) and velocities for (c))} with uniform power allocation  {across both OFDM subcarriers and symbols}, with $K=32$, $M=16$ and $K_G=8$. The index, $q$, is related to delay-Doppler bin $(n,~v)$ as follows: $q = n + (v + M/2) K_G$.  {In (b) and (c), we see that CSs with varying magnitudes as well as $v_\epsilon \neq 0$ lead to AF sidelobes, which would increase the OP.}}}
\label{fig_histogram_cross_ambiguity}
\end{figure}

\begin{nrem}
\label{rem:AF}
$p_{k,m}$ in (\ref{eq_ML_reformulate_define2}) denotes the CS power at the $k^{\mathrm{th}}$ OFDM subcarrier of the $m^{\mathrm{th}}$  OFDM symbol. $\gamma_{n,v,v_{\epsilon}}$ in \eqref{eq_ML_reformulate_define3} represents the cross-correlation between the transmitted signals and the received distorted signals, and $r_{n,v,v_{\epsilon}}$ in \eqref{eq_ML_reformulate_define} represents the discrete AF of the transmitted signals.  {For $r_{n,v,v_{\epsilon}}$ in \eqref{eq_ML_reformulate_define}, the random generalization of CSs gives random $p_{k,m}$ and induces sidebobes to $r_{n,v,v_{\epsilon}}$. Also, the fractional Doppler, $v_\epsilon$, yields an $M$-DFT evaluated at a discrete frequency that is \emph{not} an integer multiple of $1/M$. This induces sidelobes over all $M$ Doppler bins to $r_{n,v,v_{\epsilon}}$ in general. }  As an illumination, Fig. ~\ref{fig_histogram_cross_ambiguity} plots $|r_{n,v,v_{\epsilon}}|$ and $|\gamma_{n,v,v_{\epsilon}}|$ for: (a) BPSK input (i.e., constant magnitude CSs) and $v_{{\epsilon}}=0$, (b)  {256-QAM} input (i.e., variable magnitude CSs) and $v_{{\epsilon}}=0$, and (c) BPSK input and $v_{{\epsilon}}\neq 0$. To successfully locate $(n_0,~v_0)$ from $|\gamma_{n,v,v_{\epsilon}}|$, it is desirable for the peak lobe at $|r_{0,0,v_{\epsilon}}|$ to be easily distinguishable from its sidelobes.   {Fig. \ref{fig_histogram_cross_ambiguity}b (Fig. \ref{fig_histogram_cross_ambiguity}c) show that CSs with variable magnitudes (non-zero $v_{\epsilon}$) lead to AF sidelobes, which would increase the probability of incorrectly estimating $(n_0,~ v_0)$, i.e., the OP.}  {As seen in Fig. \ref{fig_histogram_cross_ambiguity}, the sidelobe levels are a function of both the CS magnitudes and $v_\epsilon$. Hence, the overall sidelobe level can be suppressed (and along with it the OP) through the careful design of CS input distributions (which is under our control), accounting for the effect of $v_\epsilon$. This motivates the problem of minimizing the OP w.r.t the input distribution.
}
\end{nrem}

\subsection{ML-based radar metric}
\label{subsec:ML-based radar metric}
A metric that directly characterizes the ML estimator's sensing performance is the OP -- the probability of incorrectly estimating $(n_0,~v_0)$. 

\begin{figure}[t]
\centering
\includegraphics[width=2.2in]{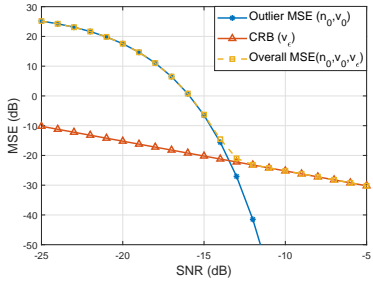}
\caption{\textcolor{blue}{The overall MSE, the outlier MSE and the CRB relationship for $K=32$, $K_G=8$, $M=8$.}}
\label{fig_appr_MSE}
\end{figure}

\begin{nrem}
\label{rem:CRB}
 {The popular CRB is not adopted as the radar metric. As a reason, the overall MSE of $(n_0,~v_0,~v_{\epsilon})$ in \eqref{eq_ML0} is composed of: $ \mathrm{MSE}\left( n,~v,~v_{\epsilon}\right)= \mathrm{MSE}\left( n,~v\right) + \mathrm{MSE}\left( v_{\epsilon}\right)$. The first term on the right-hand side, $\mathrm{MSE}\left( n,~v\right)$, captures the contribution to the MSE from incorrectly estimating the delay-Doppler bin $(n_0,~v_0)$, and is known as the outlier MSE. The second term captures the MSE w.r.t $v_\epsilon$, conditioned on having correctly estimated the target's delay-Doppler bin $(n_0,~v_0)$, and is bounded below by the CRB. \textcolor{blue}{Fig. \ref{fig_appr_MSE} illuminates the above relationship with theoretical approximations \cite{athley2005threshold}.} For the on-grid estimation in our scenario, the outlier MSE should be our objective, and the outliers, from Fig. \ref{fig_appr_MSE}, mostly occur at the threshold SNR region \footnote{\textcolor{blue}{The threshold region is a typical radar phenomenon and refers to the SNR region where radar's OP/MSE decreases sharply from a saturating high value to a saturating low value, which is the region of interest. Beyond the threshold region, the MSE usually approaches the CRB.}} and below \footnote{Hence, in the DFRC scenario with variable magnitude input distributions in Section \ref{sec_Gaussian_opt}, the average UBOP is approximated in the low SNR region.}. On the contrary, the overall MSE only converges to the CRB in the high SNR region when $v_\epsilon$ is also the estimation objective. Moreover, in our on-grid delay-Doppler bin estimation, the CRB is not also suitable because the parameters are discrete, breaking the regularity condition on which the CRB is defined \cite{kay1993fundamentals}. \textcolor{blue}{Therefore, in this paper, we focus on mitigating outliers in on-grid Delay-Doppler estimation and use OP as the analytical radar metric, which is a scaled version of the outlier MSE but offers a simpler expression. } }
\end{nrem}

\begin{figure}[t]
\centering
\includegraphics[width=2.2in]{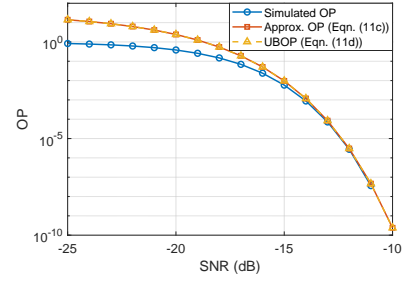}
\caption{\textcolor{blue}{The simulated and approximated OP for $K=32$, $K_G=8$, $M=8$.}}
\label{fig_appr_L8}
\end{figure}

To account for the  {arbitrariness} of $v_{{\epsilon}}$, we assume it is uniformly distributed over $(-1/2,~1/2)$, and consider the OP averaged over $v_{\epsilon}$. This quantity, denoted by $\nbbP_{\mathrm{o}}$, is the probability that the  {sidelobe magnitudes ($|\gamma_{n,v,v_{\epsilon}}|$ for $(n,~v)\neq (0,~0)$)} is larger than the peak lobe magnitude ($|\gamma_{0,0,v_{\epsilon}}|$), which can be expressed as follows
\begin{subequations}
\begin{align}
\label{eq_prob_outlier}
\nbbP_{\rm o} &\triangleq  \mathbb{E}_{v_{{\epsilon}}}\Big\{\nbbP\Big[\bigcup_{n,v}\left\lbrace |\gamma_{n,v,v_{\epsilon}}|>|\gamma_{0,0,v_{\epsilon}}|\Big\}\Big]\right\rbrace \\
\label{ineq:ub}
&\leq \mathbb{E}_{v_{{\epsilon}}}\Big\{\sum_{
(n,~v)\neq (0,~0)}\nbbP\left[  |\gamma_{n,v,v_{\epsilon}}|>|\gamma_{0,0,v_{\epsilon}}|\right]\Big\}\\
\label{eq_prob_outlier2}
&\approx \mathbb{E}_{v_{{\epsilon}}}\Big\{\sum_{
(n,~v)\neq (0,~0)}\exp\Big\{-\frac{|r_{0,0,v_{\epsilon}}|}{2\sigma^2}\Big\}I_0\left(\frac{|r_{n,v,v_{\epsilon}}|}{2\sigma^2}\right)/2\Big\}\\\label{eq_prob_outlier3}
&\leq \mathbb{E}_{v_{{\epsilon}}}\Big\{\sum_{
(n,~v)\neq (0,~0)}\exp\Big\{-\frac{|r_{0,0,v_{\epsilon}}|-|r_{n,v,v_{\epsilon}}|}{2\sigma^2}\Big\}/2\Big\} \\
&\overset{\triangle}{=} \nbbP_{\rm UB},
\end{align} 
\end{subequations}
where $\sigma^2=\sigma_{\mathrm{N}}^2/A^2$. The inequality in \eqref{ineq:ub} stems from the union bound, while the upper bound in \eqref{eq_prob_outlier3}, termed the UBOP, is more tractable for optimization w.r.t the input distribution than $\nbbP_\mathrm{o}$. Hence, we adopt the UBOP as the radar metric from hereon.  {The derivation for \eqref{eq_prob_outlier2}-\eqref{eq_prob_outlier3} is provided in Appendix \ref{sec_App_OP} and is shown to capture the simulated OP soundly in Fig. ~\ref{fig_appr_L8}.}

\begin{nrem}\label{Re:OP_AF}
 {
The expression in \eqref{eq_prob_outlier3} captures the relationship between the UBOP and the AF \cite{1597550, moulin2022joint, chen2018waveform}. Specifically, at low SNR, $\sum_{
(n,~v)\neq (0,~0)}\exp\left\lbrace-{(|r_{0,0,v_{\epsilon}}-r_{n,v,v_{\epsilon}})|}/{2\sigma^2}\right\rbrace \propto\sum_{
(n,~v)\neq (0,~0)}-{(|r_{0,0,v_{\epsilon}}|-|r_{n,v,v_{\epsilon}}|)}/{2\sigma^2}$,  indicating that the UBOP is approximately inversely proportional to the integrated-side-to-peak-lobe-difference (ISPLD) of the AF in this regime. On the contrary, at high SNR, the maximal exponential term in \eqref{eq_prob_outlier3} dominates. Hence, the UBOP in this regime is inversely proportional to the peak-to-the-largest-side-lobe difference of the AF. The relationship with the AF further demonstrates that the OP and UBOP are sound and well-motivated metrics.}
\end{nrem}

\begin{nrem} 
 {From \eqref{eq_prob_outlier3} (combining with \eqref{eq_ML_reformulate_define}), the UBOP is a function of the input distribution through $p_{k,m}$, which is only related with the CS magnitudes. This motivates us to consider the problem of minimizing the UBOP under two broad classes of input distributions, (i) those with constant magnitudes (e.g., PSK input distribution), and (ii) those with variable magnitudes (e.g., Gaussian input distribution), which we consider in the following sections. }
\end{nrem}

\section{Minimizing OP for Constant Magnitude Input Distributions}
\label{sec_BPSK}
In this section, we minimize the UBOP ($\nbbP_{\rm UB}$ in \eqref{eq_prob_outlier3}) for CSs with constant magnitude input distributions, such as PSK. In this scheme, the UBOP is deterministic regardless of the random information carried by CSs' phases. Hence, waveforms with constant magnitude input distributions are able to achieve the same radar performance in DFRC as that of using deterministic OFDM  {waveforms} in a pure radar system.

\subsection{Problem formulation}
 \textcolor{blue}{For the special case of constant magnitude CS, the design of its input distribution includes two parts, (i) the power allocation across OFDM subcarriers and OFDM symbols, i.e., $p_{k,m}$, and (ii) the probability mass vector (PMV) of each constellation point. In this context, since the UBOP is a deterministic function of $p_{k,m}$ and is independent of the PMV, we formulate the following optimization problem}
\begin{mini!}
    {\{p_{k,m}\}}{
\nbbP_{\rm UB}\text{ in }\eqref{eq_prob_outlier3},}{\label{eq_optimization_P1}}{\label{eq_optimization_P1_1}}
       \addConstraint{\sum_{k,~m} p_{k,m} \leq  P^{\max}}\label{eq_optimization_P1_3}
        \addConstraint{
        p_{k,m} \geq 0, ~\forall k,~m,}\label{eq_optimization_P1_3new}
\end{mini!}
where $P^{\max}$ is the transmit power constraint. 

\begin{nrem}
    \textcolor{blue}{In this section, we primarily focus on optimizing radar performance, while still acknowledging that information can be conveyed by phases. Hence, we implicitly assume a uniform PMV over the constellation symbols which can achieve a rate of up to $\log_2 Q$  bits/symbol for $Q-$PSK \cite{1188416}, and disregard the communication constraint in \eqref{eq_optimization_P1}. Consequently,} this scheme of constant magnitude input distributions provides a performance benchmark for the scheme of variable magnitude input distributions (Gaussian input) in Section \ref{sector_Gaussian}. This part will also inspire the simplified algorithm for the Gaussian input scheme in section \ref{sec_Gaussian_decouple}. 
\end{nrem}

For the above problem, we have the following result:
\begin{lem}\label{lem1}
The optimal power allocation across subcarriers ($p_{k,m} $ across $k$ for $\forall~m$) to minimize $\nbbP_{\rm UB}$ is the uniform power allocation.
\end{lem}
\begin{IEEEproof} 
Let $p_{m} \triangleq \sum_{k} ~p_{k,m}$ denote the total power of the $m^{\mathrm{th}}$ OFDM symbol. The expectation over $v_{\epsilon}$ in \eqref{eq_prob_outlier3} is approximated by taking multiple samples of $v_{{\epsilon}}$ from the set $\{{\epsilon}_1, ~\cdots, ~{\epsilon }_{|\kappa|}\}$ (denoted by set $\kappa$, the larger the number of samples, the more accurate the approximation)  following its uniform distribution. Hence, $\nbbP_{\rm UB}$ becomes:
\begin{subequations}
\begin{align}
\nonumber\label{eq_UB_uniform}
&\nbbP_{\rm UB}\\
\approx &\frac{1}{\vert\kappa\vert}\sum_{{{\epsilon}}\in \kappa}\sum_{(n,~v)\neq (0,~0)}\exp\Big\{-\frac{|r_{0,0,{\epsilon}}|-|r_{n,v,{\epsilon}}|}{2\sigma^2}\Big\} \hspace{3mm}(\mbox{for large }|\kappa|)\\\nonumber\label{eq_UB_uniform10} 
=& \frac{1}{\vert\kappa\vert}\sum_{{{\epsilon}}\in \kappa} \sum_{n=0,~v\neq 0}\exp\Big\{-\frac{|r_{0,0,{\epsilon}}|-|r_{n,v,{\epsilon}}|}{2\sigma^2}\Big\}\\&+\frac{1}{\vert\kappa\vert}\sum_{{{\epsilon}}\in \kappa}\sum_{~n\neq 0}\exp\Big\{-\frac{|r_{0,0,{\epsilon}}|-|r_{n,v,{\epsilon}}|}{2\sigma^2}\Big\}
\\\label{eq_UB_uniform1}
\geq&  \frac{\sum_{{{\epsilon}}\in \kappa}}{\vert\kappa\vert}\Big(\sum_{v\neq 0}\exp\Big\{-\frac{|r_{0,0,{\epsilon}}|-|r_{0,v,{\epsilon}}|}{2\sigma^2}\Big\}+\sum_{n\neq 0}\exp\Big\{-\frac{|r_{0,0,{\epsilon}}|}{2\sigma^2}\Big\}\Big)\\\nonumber\label{eq_UB_uniform2}
=& \frac{\mathbf{1}^T}{\vert\kappa\vert}\exp\Big\{-\frac{{\mathbf{1}_{M-1}}\otimes|\mathbf{r}_0 |-|\mathbf{r'}| }{2\sigma^2}\Big\}+\frac{(K_G-1)M}{\vert\kappa\vert}\mathbf{1}^T\exp\Big\{-\frac{|\mathbf{r}_0| }{2\sigma^2}\Big\}\\
\overset{\triangle}{=} &\nbbP_{\rm UB}(\mathbf{r}),
\end{align}
\mbox{where}
\begin{align}
 \label{eq_vector_r_v}
\mathbf{r}_{v}=&\left[r_{0,v,{\epsilon}_1}, ~\cdots, ~r_{0,v,{\epsilon}_{|\kappa|}}\right]^T\in\mathbb{C}^{|\kappa|}
,\\
\mathbf{r'}=&\left[\mathbf{r}_{1}^T, ~\cdots, ~\mathbf{r}_{M-1}^T\right]^T\in\mathbb{C}^{(M-1)|\kappa|},
\\\mathbf{r}=&\left[\mathbf{r}_{0}^T,~\mathbf{r'}^T\right]^T=\mathbf{F}^\mathrm{D}\mathbf{p} \in\mathbb{C}^{M|\kappa|},\\
\mathbf{p} =&\left[p_{0}, ~\cdots, ~p_{M-1}\right]^T\in\mathbb{R}^{M},\\
\mathbf{F}^{\mathrm{D}}=&\Big[\mathbf{f}^{\mathrm{D}}_{1-M/2,{\epsilon}_1},~\mathbf{f}^{\mathrm{D}}_{1-{M}/2,{\epsilon}_{2}},~\cdots,~\mathbf{f}^{\mathrm{D}}_{M/2-1,{\epsilon}_{|\kappa|}}\Big]^T\in\mathbb{C}^{M|\kappa|\times M},\\\label{eq_define_F_D}
{f}^{\mathrm{D}}_{v,{\epsilon}_l,m}=&e^{-j2\pi{m(v-{{\epsilon}_l})}/{M}},~l=(1,~\cdots,~|\kappa|),
\end{align}
\end{subequations}
where \eqref{eq_UB_uniform1} is achieved because $|r_{n,v,{\epsilon}}|\geq 0$ for $n\neq 0$ with the equality  achieved if and only if $p_{k,m}=p_{m}/K$ since $r_{n,v,{\epsilon}}=\sum_{m,k} p_{m}e^{j2\pi\frac{nk}{K}}e^{-j2\pi \frac{v m} {M}}e^{j2\pi \frac{{\epsilon}m} {M}}/K=\sum_{k} e^{j2\pi\frac{nk}{K}}\sum_m p_{m}e^{-j2\pi \frac{v m} {M}}e^{j2\pi \frac{{{\epsilon}}m} {M}}/K=0$. Then in \eqref{eq_UB_uniform1}, the only variables involved are the power allocation across OFDM symbols, i.e., $p_{m}$  ($r_{0,v,{\epsilon}}=\sum_m p_{m}e^{-j2\pi \frac{v m} {M}}e^{j2\pi \frac{{{\epsilon }}m} {M}}$). Hence, the optimal power allocation would have uniform power allocation across OFDM subcarriers, and we need to optimize the power allocation across OFDM symbols, i.e., $p_{m}$ for $\forall m$ or equivalently $\mathbf{p}$.
\end{IEEEproof}

With uniform (optimal) power allocation across OFDM subcarriers, the optimization problem \eqref{eq_optimization_P1} is simplified into:
\begin{mini!}
    {\mathbf{p} ,~\mathbf{r}}{\nbbP_{\rm UB}(\mathbf{r})\text{ in }\eqref{eq_UB_uniform2},}{\label{eq_optimization_P2}}{\label{eq_optimization_P2_1}}
       \addConstraint{\mathbf{1}^T \mathbf{p}  \leq  P^{\max}}\label{eq_optimization_P2_2}
         \addConstraint{ \mathbf{p}\succeq \mathbf{0} }\label{eq_optimization_P2_2new}
         \addConstraint{\mathbf{r}=\mathbf{F}^{\mathrm{D}}\mathbf{p} .}\label{eq_optimization_P2_3}
\end{mini!}
 {After obtaining the optimized power allocation across the OFDM symbols from \eqref{eq_optimization_P2}, i.e. $\mathbf{p}$, the optimal power allocation at each OFDM symbol's sub-carrier is $p_{k,m}=p_m/K$, $\forall k,~m$ (from Lemma \ref{lem1}).}

\subsection{Low complexity optimization}
Solving the non-convex problem \eqref{eq_optimization_P2} can be very time-consuming with a large number of OFDM symbols $M$, whereas in OFDM radar, $M$ shall be large, i.e., above $512$, to achieve high Doppler resolution \cite{sturm2011waveform}. {Hence, in this section, we adopt a low-complexity alternating direction method of multipliers (ADMM) algorithm where problem \eqref{eq_optimization_P2} is approximately solved in an iterative manner with closed-form solutions in each round.}

Considering the linear constraint in \eqref{eq_optimization_P2_3}, we adopt ADMM to iteratively update $\mathbf{p} $ and $\mathbf{r}$ until the two successive solutions get close \cite{bertsekas2014constrained}. Following the ADMM structure\footnote{With ADMM, we obtain  local optimal solutions (referred to as optimal in the following) as an approximation to the global optimal solution in this paper, similarly to the BCD algorithm in Section \ref{sec_Gaussian_opt}.}, at the $l^{\mathrm{th}}$ iteration,  problem \eqref{eq_optimization_P2} is iteratively formulated as:
\begin{align}
\label{eq_ADMM_r}
\mathbf{r}^{(l+1)} &:= \underset{\mathbf{r}}{\arg\min} \left\lbrace \nbbP_{\rm UB}(\mathbf{r})+\frac{\rho}{2}\|\mathbf{r}-\mathbf{F}^{\mathrm{D}}\mathbf{p} ^{(l)}+\mathbf{d}^{(l)}\|^2\right\rbrace, \\\label{eq_ADMM_p}
\mathbf{p} ^{(l+1)} &:=\underset{\mathbf{p} }{\arg\min}\left\lbrace \mathbbm{1}_{(\mathbf{1}^T\mathbf{p}  \leq P^{\max},~\mathbf{p}\succeq \mathbf{0})}+\frac{\rho}{2}\|\mathbf{r}^{(l+1)}-\mathbf{F}^{\mathrm{D}}\mathbf{p} +\mathbf{d}^{(l)}\|^2\right\rbrace,
\\\label{eq_ADMM_d}
\mathbf{d}^{(l+1)} &:=\mathbf{d}^{(l)} +\mathbf{r}^{(l+1)}-\mathbf{F}^{\mathrm{D}}\mathbf{p} ^{(l+1)},
\end{align}
where $\mathbf{d}^{(l)}$ is the ADMM stepsize at the $l^{\mathrm{th}}$ iteration and $\rho$  is the ADMM penalty coefficient. 

\subsubsection{ {Updating $\mathbf{r}$ in \eqref{eq_ADMM_r}}}
To solve subproblem \eqref{eq_ADMM_r}, we adopt the successive convex programming (SCP) to approximate the exponential function in \eqref{eq_ADMM_r} linearly by the first-order Taylor expansion. The optimal solution of this approximated problem is then used as a new operating point of the Taylor expansion for the next SCP round. The procedure is repeated until two successive solutions are close enough. 

Consequently, assume that $\mathbf{r}^{(l+1,~t)}$ is the optimal solution at the $t^{\mathrm{th}}$ iteration of SCP. Then $\mathbf{r}^{(l+1,~t+1)}$ is updated as:
\begin{subequations}
\begin{align}
\mathbf{r}^{(l+1,~t+1)}&=\underset{\mathbf{r}}{\arg\min }\left\lbrace{\boldsymbol{\alpha}^{(l+1,~t)}}^T|\mathbf{r}|+\frac{\rho}{2}\|\mathbf{r}-\mathbf{F}^{\mathrm{D}}\mathbf{p} ^{(l)}\|^2 \right\rbrace\\
\label{eq_BCD_r3}
&=\left(\mathbf{F}^{\mathrm{D}}\mathbf{p} ^{(l)}- {\boldsymbol{\theta}^{(l)}\odot\boldsymbol{\alpha}^{(l+1,~t)}} /\rho\right)^+,
\end{align}
\end{subequations}
where $\boldsymbol{\theta}^{(l)}=\angle\mathbf{F}^{\mathrm{D}}\mathbf{p} ^{(l)}$. The closed-form expression in \eqref{eq_BCD_r3} is derived in Appendix \ref{sec_APP_OPT}. The first-order Taylor coefficient $\boldsymbol{\alpha}^{(l+1,~t)}=[{\boldsymbol{\alpha}_0^{(l+1,~t)}}^T, ~\cdots, ~{\boldsymbol{\alpha}_{M-1}^{(l+1,~t)}}^T]^T$ is expressed in \eqref{eq_ADMM_r_Coeff}.
\begin{figure*}
\begin{equation}
\label{eq_ADMM_r_Coeff}
\boldsymbol{\alpha}_v^{(l+1,t)}=\begin{cases}-\sum_{v=1}^{M-1}\exp\Big\{-\frac{|{\mathbf{r}_0^{(l+1,t)}}|-|{\mathbf{r}_v^{(l+1,t)}}|}{2\sigma^2}\Big\}/{\left(2\sigma^2{\vert\kappa\vert}\right)}-(K_G-1)M\exp\Big\{-\frac{|{\mathbf{r}_0^{(l+1,t)}}|}{2\sigma^2}\Big\}/{\left(2\sigma^2{\vert\kappa\vert}\right)},&\text{ for }v=0;\\ \exp\Big\{-\frac{|{\mathbf{r}_0^{(l+1,t)}}|-|{\mathbf{r}_v^{(l+1,t)}}|}{2\sigma^2}\Big\}/{\left(2\sigma^2{\vert\kappa\vert}\right)},&\text{ otherwise.}\end{cases}
\end{equation}
\noindent\makebox[\linewidth]{\rule{\paperwidth}{0.4pt}}
\end{figure*}

\subsubsection{ {Updating $\mathbf{p} $ in \eqref{eq_ADMM_p}}}
Given a fixed $\mathbf{r}^{(l+1)}$, the optimal $\mathbf{p} ^{(l+1)}$ can be directly solved by quadratic programming (QP) as:
\begin{mini!}
    {\mathbf{p} }{\frac{\rho}{2}\|\mathbf{r}^{(l+1)}-\mathbf{F}^\mathrm{D}\mathbf{p} \|^2,}{\label{eq_optimization_P22}}{\label{eq_optimization_P21_1}}
       \addConstraint{\mathbf{1}^T \mathbf{p}  \leq  P^{\max}}\label{eq_optimization_P22_2}
       \addConstraint{ \mathbf{p}\succeq \mathbf{0} .}\label{eq_optimization_P2_2new}
\end{mini!}
The algorithm is summarized in Algorithm \ref{ADMM}.
\begin{algorithm}[t] 
 $\textbf{Input}$: $l\leftarrow 0, ~\mathbf{r}^{(0)},~\mathbf{p} ^{(0)}, ~ \rho, ~ {\epsilon}_{S}>0$\\
 $\textbf{Repeat (Solving $\mathbf{p} $ in \eqref{eq_optimization_P2})}$: \\
 \begin{enumerate}
   \item Update $\mathbf{r}^{(l+1)}$ in problem \eqref{eq_ADMM_r} by SCP
  \begin{enumerate}
        \item[]  Initialize $t\leftarrow 0$; $\mathbf{r}^{(l+1,~0)}\leftarrow \mathbf{r}^{(l)}$
        \item[] \textbf{while $t\leq 10^2$} 
        \begin{enumerate}
            
            \item[]  Compute $\boldsymbol{\alpha}^{(l+1,~t)} $ in \eqref{eq_ADMM_r_Coeff} at the operating point $\mathbf{r}^{(l+1,~t)}$

             \item[]  Compute $\mathbf{r}^{(l+1,~t+1)}$ using \eqref{eq_BCD_r3}

             \item[]  $t\leftarrow t+1$; $\mathbf{r}^{(l+1)} \leftarrow \mathbf{r}^{(l+1,~t+1)}$

         \end{enumerate}
          \item[] Iteration complexity $\mathcal{O}\left(M^2|\kappa| \right)$
     \end{enumerate}
   \item Update $\mathbf{p} ^{(l+1)}$ in \eqref{eq_ADMM_p} by QP (complexity $\mathcal{O}\left(M^3 \right)$)

   \item $l\leftarrow l+1$; ${\mathbf{p}} \leftarrow \mathbf{p} ^{(l+1)}$

   \item Quit if $\| \mathbf{p} ^{(l+1)}-\mathbf{p} ^{(l)}\|< {\epsilon}_{S}\|\mathbf{p} ^{(l+1)}\|$
 \end{enumerate}
  $\textbf{Output}$: ${\mathbf{p}}_{k,m}\leftarrow p_{m}/K$ \\
\caption{ {ADMM for PSK input in \eqref{eq_optimization_P1}}}
\label{ADMM}
\end{algorithm}

\section{Minimizing OP for variable magnitude Input Distributions (Gaussian)}
\label{sector_Gaussian}
This section evaluates the radar performance in OFDM DFRC with non-zero mean Gaussian input, where the magnitude of the CSs becomes a random variable obeying the Rician distribution. Consequently, we evaluate the aUBOP and then optimize the Gaussian mean and variance at each symbol's subcarriers to minimize the aUBOP. In this scheme, an achievable communication constraint is added to avoid a zero-variance optimization result where no information bits are transmitted.

\subsection{Problem formulation}
In this context, we first model the CS $X[k,~m]$ in \eqref{eq_CP_discrete} as a complex non-zero mean asymmetric Gaussian variable:
\begin{align}\label{eq_X_ID}
X[k,~m]={X}_{\mathrm{R}}[k,~m]+jX_{\mathrm{I}}[k,~m],
\end{align}
where ${X}_{\mathrm{R/I}}[k,~m]\sim\mathcal{N}({\mu}_{\mathrm{R/I},~k,m},~{\sigma}_{\mathrm{R/I},~k,m}^2)$.

Then,  the average power (second-order moments) allocated to the real and imaginary part of the $m^{\mathrm{th}}$ OFDM symbol at the $k^{\mathrm{th}}$ subcarrier are defined as:
%
\begin{subequations}\label{eq_Gaussian_power_define}
\begin{align}
\overline{p}_{\mathrm{R},k,m}&={\mu}_{\mathrm{R},{k,m}}^2+{\sigma}_{\mathrm{R},{k,m}}^2,\\
\overline{p}_{\mathrm{I},k,m}&={\mu}_{\mathrm{I},{k,m}}^2+{\sigma}_{\mathrm{I},{k,m}}^2.
\end{align}   
\end{subequations}

On this basis, the AF $r_{n,v,v_{\epsilon}}$ in \eqref{eq_prob_outlier3} becomes random, making the UBOP varying. Hence, we derive the aUBOP as follows  {(abbreviating the expectation over $X[k,~m]$ as $X$, i.e., $\mathbb{E}_{X}\triangleq\mathbb{E}_{X[k,~m]}$ )}
\begin{subequations}
\begin{align}
\label{eq_averaged_outlier}
&\overline{\nbbP}_{\rm UB}\left(\overline{\mathbf{p}},~\boldsymbol{\sigma}\right)\\
=&\mathbb{E}_{v_{{\epsilon}},X}\Big\{\sum_{
(n,~v)\neq (0,~0)}\exp\Big\{-\frac{|r_{0,0,v_{\epsilon}}|-|r_{n,v,v_{\epsilon}}|}{2\sigma^2}\Big\}/2 \Big\}\\\label{eq_averaged_outlier1}
\propto& \frac{1}{4\sigma^2}\sum_{(n,~v)\neq (0,~0)}\mathbb{E}_{{{\epsilon}},X}\left\lbrace-{|r_{0,0,{\epsilon}}|+|r_{n,v,{\epsilon}}|}\right\rbrace\\\label{eq_averaged_outlier2}
\leq&\frac{1}{4\sigma^2}\sum_{
(n,~v)\neq (0,~0)}\mathbb{E}_{ {{\epsilon}}}\Big\{-|\mathbb{E}_{X}\left\lbrace r_{0,0,{\epsilon}}\right\rbrace|+\sqrt{\mathbb{E}_{X}\left\lbrace|r_{n,v,{\epsilon}}|^2\right\rbrace}\Big\}\\\label{eq_averaged_outlier3}
= &\frac{1}{4|\kappa|\sigma^2}\sum_{{\epsilon}\in\kappa}\sum_{(n,~v)\neq (0,~0)}-\sqrt{\overline{\mathbf{p}}^T\mathbf{A}_{{\epsilon}}\overline{\mathbf{p}} }+g_{n,v,{\epsilon}}\left(\overline{\mathbf{p}},~\boldsymbol{\sigma}\right),
\end{align}
\begin{align}\label{eq_g_nv}
\mbox{with}~  
g_{n,v,{\epsilon}}\left(\overline{\mathbf{p}},~\boldsymbol{\sigma}\right)
=&\sqrt{\mathbf{\overline{p}}^T\mathbf{A}_{n,v,{\epsilon}}\mathbf{\overline{p}}-2\|\overline{\mathbf{p}}-\boldsymbol{\sigma}\|^2},
\end{align}
\begin{align}
\setlength{\jot}{-5pt}
\overline{\mathbf{p}}=&\left[\overline{p}_{\mathrm{R},{0,0}}, ~ \overline{p}_{\mathrm{R},{1,0}},~\cdots,,~ \overline{p}_{\mathrm{I},{0,0}},~ \cdots,~ \overline{p}_{\mathrm{I},{K-1,M-1}}\right]^T,\\
 {\boldsymbol{\sigma}}=&\left[ {\sigma}_{\mathrm{R},{0,0}}, ~ {\sigma}_{\mathrm{R},{1,0}},~\cdots,~ {\sigma}_{\mathrm{I},{0,0}},~ \cdots,~{\sigma}_{\mathrm{I},{K-1,M-1}}\right]^T,\\
\label{eq_g_nv2}
\mathbf{A}_{{\epsilon}}
=&\mathfrak{R}\left\lbrace{\mathbf{f}_{0,0,{{\epsilon}}}}\right\rbrace\mathfrak{R}\left\lbrace{\mathbf{f}_{0,0,{{\epsilon}}}}\right\rbrace^T+\mathfrak{I}\left\lbrace{\mathbf{f}_{0,0,{{\epsilon}}}}\right\rbrace\mathfrak{I}\left\lbrace{\mathbf{f}_{0,0,{{\epsilon}}}}\right\rbrace^T,\\\label{eq_g_nv3}
\mathbf{A}_{n,v,{\epsilon}}=&{\mathfrak{R}\left\lbrace\mathbf{f}_{n,v,{\epsilon}}\right\rbrace}{\mathfrak{R}\left\lbrace\mathbf{f}_{n,v,{\epsilon}}\right\rbrace}^T+{\mathfrak{I}\left\lbrace\mathbf{f}_{n,v,{\epsilon}}\right\rbrace}{\mathfrak{I}\left\lbrace\mathbf{f}_{n,v,{\epsilon}}\right\rbrace}^T+2\mathbf{I}_{2KM}, \\
\mathbf{f}_{n,v,{{\epsilon}}}=&\mathbf{1}_2\otimes\left(\mathbf{f}^{\mathrm{D}}_{v,~{{\epsilon}}}\otimes \mathbf{f}^{\mathrm{R}}_{n}\right)\in\mathbb{C}^{2KM},\\
{f}^{\mathrm{R}}_{n,k}=&e^{j2\pi{nk}/{K}},
\end{align}
\end{subequations}
where \eqref{eq_averaged_outlier1} assumes low SNR at the radar's receiver\footnote{ {
In this paper, we only discuss the statistic approximation and the optimization in the low SNR region due to the space limit, while similar procedures can be done for the middle SNR and for the high SNR scenarios. For the middle SNR region, \eqref{eq_averaged_outlier1} can be replaced by Taylor approximation with appropriate orders and coefficients given its finite SNR  region\cite{fraser1965survey}. For the high SNR region, inspired by Remark \ref{Re:OP_AF}, the objective function might be approximated by maximizing the minimal average peak-to-side-lobe difference.}}, giving $\exp\{x\}\rightarrow x+1$ for $x\rightarrow 0 $ ($r_{0,0,\epsilon}\geq r_{n,v,\epsilon}$ for any $(n,~v)$). \eqref{eq_averaged_outlier2} is due to the concavity of the square root function. \eqref{eq_averaged_outlier3} is achieved from Appendix \ref{sec_App_2nd_monents},  {whose first sum term refers to $\mathbb{E}_{ {{\epsilon}}}\left\lbrace-|\mathbb{E}_{X}\left\lbrace r_{0,0,{\epsilon}}\right\rbrace|\right\rbrace$ in \eqref{eq_averaged_outlier2}  representing the magnitude of average peak lobe, and whose second sum term refers to $\mathbb{E}_{ {{\epsilon }}}\left\lbrace-|\mathbb{E}_{X }\left\lbrace r_{n,v,{\epsilon}}\right\rbrace|\right\rbrace$  at $(n,~v)\neq~(0,~0)$ in \eqref{eq_averaged_outlier2} representing the magnitude of the average sidelobes}.

\begin{nrem}
\eqref{eq_averaged_outlier2} indicates that the optimization objective (aUBOP), $\overline{\nbbP}_{\rm UB}\left(\overline{\mathbf{p}},~\boldsymbol{\sigma}\right)$, finally becomes a scaling term of the average ISPLD in low SNR, corresponding to \textit{Remark} \ref{Re:OP_AF}.  {Also, \eqref{eq_averaged_outlier3} indicates that, without communication constraints, the optimal solution to minimize \eqref{eq_averaged_outlier3} yields maximizing $\|\mathbf{\overline{p}}-\boldsymbol{\sigma}\|$ for an arbitrary $\mathbf{\overline{p}}$, i.e., $\sigma^2_{\mathrm{R/I},k,m}=0$ and hence $\overline{p}_{\mathrm{R/I},k, m}=\mu_{\mathrm{R/I},k, m}^2$. In this context, all the power is allocated to the deterministic component of the Gaussian distributions for the radar-optimal purpose. In other words, the randomness of the CS  magnitudes degrades radar's OP.}
\end{nrem}

\subsection{Optimization}
\label{sec_Gaussian_opt}
We then formulate the corresponding optimization problem in DFRC using the developed metric in \eqref{eq_averaged_outlier3}. With the constraint on achievable rate, we have the optimization problem (neglecting the constant multiplier $1/(4\sigma^2)$):
\begin{mini!}
    {\mathbf{\overline{p}},~\boldsymbol{\sigma}}{\sum_{{\epsilon}\in\kappa}\sum_{(n,~v)\neq (0,~0)}-\sqrt{\overline{\mathbf{p}}^T\mathbf{A}_{{\epsilon}}\overline{\mathbf{p}} }+g_{n,v,{\epsilon}}\left(\overline{\mathbf{p}},~\boldsymbol{\sigma}\right),}{\label{eq_optimization_P3}}{\label{eq_optimization_P3_1}}
       \addConstraint{{\mathbf{1}}^{T}\overline{\mathbf{p}}\leq  P^{\max}}\label{eq_optimization_P3_2}
       \addConstraint{ \frac{1}{2KM}\log~ \det \left(\mathbf{I}_{2KM}+\mathbf{C}\mathrm{diag}\{\boldsymbol{\sigma}\}\right) \geq R^{\mathrm{c}}}\label{eq_optimization_P3_32}
      \addConstraint{\overline{\mathbf{p}} \succeq \boldsymbol{\sigma}}\label{eq_optimization_P3_33}
        \addConstraint{ \boldsymbol{\sigma}\succeq \mathbf{0},}\label{eq_optimization_P3_4}
\end{mini!}
where \eqref{eq_optimization_P3_32} is the average rate constraint over $M$ OFDM symbols. In \eqref{eq_optimization_P3_32}, $\mathbf{C}=2K\mathrm{diag}\left\lbrace{\mathbf{1}_{2M}\otimes\mathbf{h}/({B\sigma_{\mathrm{C}}^2})}\right\rbrace \in \mathbb{R}^{2KM\times 2KM}$ with the $k^{\mathrm{th}}$ entry of $\mathbf{h}$ being ${h}_k={|h_{\mathrm{C},k}|^2}$ for $k=[1, ~\cdots, ~K]$. $h_{\mathrm{C},k}$ is the complex gain of the communication channel for the $k^{\mathrm{th}}$ subcarrier. $\sigma_{\mathrm{C}}^2$ is the noise power density at the communication receiver. $R^c$ is the minimum achievable rate constraint for communications. 

For efficiency, we use the method of block coordinate descent (BCD) to solve problem \eqref{eq_optimization_P3}. In BCD, given an initial point of $\overline{\mathbf{p}}^{(l)}/\boldsymbol{\sigma}^{(l)}$ at the $l^{\mathrm{th}}$ iteration, $\overline{\mathbf{p}}^{(l+1)}$ is optimized first with fixed $\boldsymbol{\sigma}^{(l)}$. Then $\boldsymbol{\sigma}^{(l+1)}$ is optimized with fixed $\overline{\mathbf{p}}^{(l+1)}$. The iteration continues until two successive solutions get close.

\subsubsection{Updating $\mathbf{\overline{p}}$}
\label{sec_opt_p}
Given $\boldsymbol{\sigma}^{(l)}$, the sub-problem for $\overline{\mathbf{p}}$ is:
\begin{mini!}
    {\overline{\mathbf{p}}}{\sum_{{\epsilon}\in\kappa}\sum_{(n,~v)\neq (0,~0)}-\sqrt{\overline{\mathbf{p}}^T\mathbf{A}_{{\epsilon}}\overline{\mathbf{p}} }+g_{n,v,{\epsilon}}\Big(\overline{\mathbf{p}},~\boldsymbol{\sigma}^{(l)}\Big),}{\label{eq_optimization_P4}}{\label{eq_optimization_P4_1}}
       \addConstraint{\mathbf{1}^{T}\overline{\mathbf{p}}\leq  P^{\max}}\label{eq_optimization_P4_2}
      \addConstraint{\overline{\mathbf{p}} \succeq \boldsymbol{\sigma}^{(l)},}\label{eq_optimization_P4_3}
\end{mini!}
whose objective function in \eqref{eq_optimization_P4_1} is non-convex. As a solution, we adopt SCP similarly to Section IV-B, where the square root function in \eqref{eq_optimization_P4_1} is approximated by its first-order Taylor expansion $\sqrt{x}\approx (x-x_0)/(2\sqrt{x_0})+\sqrt{x_0}=x/(2\sqrt{x_0})+\sqrt{x_0}/2$.


Assume $\overline{\mathbf{p}}^{(t)}$ is the operating point at the beginning of the $t^{\mathrm{th}}$ iteration when solving problem \eqref{eq_optimization_P4}. Then, \eqref{eq_optimization_P4_1} at the $t^{\mathrm{th}}$ iteration is linearly approximated as 
\begin{subequations}
\begin{align}
\nonumber\label{eq_SCP_p}
&\sum_{{\epsilon}\in\kappa}\sum_{(n,~v)\neq (0,~0)}-\sqrt{\overline{\mathbf{p}}^T\mathbf{A}_{{\epsilon}}\overline{\mathbf{p}} }+g_{n,v,{\epsilon}}\left(\overline{\mathbf{p}},~\boldsymbol{\sigma}^{(l)}\right)
\\\approx &\overline{\mathbf{p}}^T\mathbf{B}^{(l,~t)}\overline{\mathbf{p}}+{\boldsymbol{\alpha}^{(l,~t)}}^T\overline{\mathbf{p}}+{\alpha_c}^{(l,~t)},\\
\nonumber\mbox{with}~~~~~~&\\
\mathbf{B}^{(l,~t)}=&
\sum_{{\epsilon}\in\kappa}\sum_{(n,~v)\neq (0,~0)} \mathbf{A}_{n,v,{\epsilon}}/ g_{n,v,{\epsilon}}\big(\overline{\mathbf{p}}^{(t)},~\boldsymbol{\sigma}^{(l)}\big)/2\\\nonumber
=&2\|1\oslash\mathbf{g}^{(l,~t)}\|^2\mathbf{I}_{2KM}\\\nonumber&+\big(\mathfrak{R}\left\lbrace\mathbf{F}\right\rbrace\oslash\mathbf{g}^{(l,~t)}\big)\big(\mathfrak{R}\left\lbrace\mathbf{F}\right\rbrace\oslash\mathbf{g}^{(l,~t)}\big)^T\\\label{eq_matrixB1}
&+\big(\mathfrak{I}\left\lbrace\mathbf{F}\right\rbrace\oslash\mathbf{g}^{(l,~t)}\big)\big(\mathfrak{I}\left\lbrace\mathbf{F}\right\rbrace\oslash\mathbf{g}^{(l,~t)}\big)^T,\\\nonumber\label{eq_vectoralpha1}
\boldsymbol{\alpha}^{(l,~t)}=&-(K_GM-1)\sum_{{\epsilon}\in\kappa} \mathbf{A}_{{\epsilon}}\overline{\mathbf{p}}^{(t)}/\sqrt{{\mathbf{p}^{(t)}}^T\mathbf{A}_{{\epsilon}}{\mathbf{p}^{(t)}} }\\
&-4\big(\overline{\mathbf{p}}^{(t)}-\boldsymbol{\sigma}^{(t)}\big)\|1\oslash\mathbf{g}^{(l,~t)}\|^2,\\\nonumber
\mathbf{F}=&\begin{bmatrix}
\mathbf{F}_{0,-M/2},~\mathbf{F}_{1,-M/2}, ~\cdots, ~\mathbf{F}_{K_G-1,M/2-1}\end{bmatrix},\\&~(\text{w/o }\mathbf{F}_{0,0}),\\
\mathbf{F}_{n,v}=&\begin{bmatrix}\mathbf{f}_{n,v,{\epsilon}_1}, ~\cdots, ~\mathbf{f}_{n,v,{\epsilon}_{|\kappa|}}\end{bmatrix}\in\mathbb{R}^{2KM\times |\kappa|},\\
\mathbf{g}^{(l,~t)}=&\left[4{\mathbf{q}^{(t)}+8\|\overline{\mathbf{p}}^{(t)}\|^2-8\|\overline{\mathbf{p}}^{(t)}-\boldsymbol{\sigma}^{(l)}\|^2}\right]^{1/4},\\\nonumber
\mathbf{q}^{(t)}=&{\mathbf{\overline{p}}^{(t)}}^T{\mathfrak{R}\left\lbrace\mathbf{F}\right\rbrace}\odot\big({\mathbf{\overline{p}}^{(t)}}^T{\mathfrak{R}\left\lbrace\mathbf{F}\right\rbrace}\big)^*\\
&+{\mathbf{\overline{p}}^{(t)}}^T\mathfrak{I}\left\lbrace\mathbf{F}\right\rbrace\odot\big({\mathbf{\overline{p}}^{(t)}}^T\mathfrak{I}\left\lbrace\mathbf{F}\right\rbrace\big)^*,
\end{align}
\end{subequations}
 where ${\alpha_c}^{(l,~t)}$ is the Taylor constant that can be omitted.

Hence, we form the sub-problem towards solving problem \eqref{eq_optimization_P4} as:
\begin{mini!}
    {\overline{\mathbf{p}}}{\overline{\mathbf{p}}^T\mathbf{B}^{(l,~t)}\overline{\mathbf{p}}+{\boldsymbol{\alpha}^{(l,~t)}}^T\overline{\mathbf{p}},}{\label{eq_optimization_P5}}{\label{eq_optimization_P5_1}}
       \addConstraint{\text{ }\eqref{eq_optimization_P4_2}-\text{ }\eqref{eq_optimization_P4_3},}\label{eq_optimization_P5_2}
\end{mini!}
which is a QP problem.

\subsubsection{Updating $\boldsymbol{\sigma}$}
\label{sec_opt_sigma}
 When fixing $\mathbf{\overline{p}}$ as $\mathbf{\overline{p}}^{(l)}$, we observe that the objective function in problem \eqref{eq_optimization_P3} is a monotonically increasing function with $2{\mathbf{\overline{p}}^{(l)}}^{T}\boldsymbol{\sigma}-\|\boldsymbol{\sigma}\|^2$. Hence, the sub-problem for $\boldsymbol{\sigma}$ becomes
\begin{mini!}
    {{\boldsymbol{\sigma}}}{2{\mathbf{\overline{p}}^{(l)}}^{T}\boldsymbol{\sigma}-\|\boldsymbol{\sigma}\|^2,}{\label{eq_optimization_P6}}{\label{eq_optimization_P6_1}}
       \addConstraint{ \frac{1}{2KM}\log~ \det \left(\mathbf{I}_{2KM}+\mathbf{C}\mathrm{diag}\{\boldsymbol{\sigma}\}\right) \geq R^{\mathrm{c}}}\label{eq_optimization_P6_2}
      \addConstraint{\overline{\mathbf{p}}^{(l)} \succeq\boldsymbol{\sigma}}\label{eq_optimization_P6_3}
        \addConstraint{\boldsymbol{\sigma}\succeq \mathbf{0},}\label{eq_optimization_P6_4}
        \end{mini!}
        which is convex. Again, for lower complexity to handle a large number of variables, we solve problem \eqref{eq_optimization_P6} iteratively by SCP with closed-form solutions. At the $t^{\mathrm{th}}$ iteration of the SCP sub-problem, the concave objective function is approximated by the first order Taylor expansion, and the sub-problem becomes     
   \begin{mini!}
    {{\boldsymbol{\sigma}}}{{\boldsymbol{\beta}^{(l,~t)}}^{T}\boldsymbol{\sigma},}{\label{eq_optimization_P62}}{\label{eq_optimization_P62_1}}
       \addConstraint{ \eqref{eq_optimization_P6_2}-~ \eqref{eq_optimization_P6_4}, }\label{eq_optimization_P62_2}
        \end{mini!} 
with $\boldsymbol{\beta}^{(l,~t)}=2\left(\mathbf{\overline{p}}^{(l)}-\boldsymbol{\sigma}^{(t)}\right)$. The closed-form solution for problem \eqref{eq_optimization_P62}, after KKT, is given  by
\begin{align} 
 \label{eq_OP_CF_62} 
{\sigma}_{q}=\begin{cases}{\overline{p}}^{(l)}_q, & ~ u>\frac{2KM{\beta}^{(l,~t)}_q\left(1+C_{q,q}{{\overline{p}}}^{(l)}_{q}\right)}{ C_{q,q}};\\
0, &~ 0< u <\frac{2KM{\beta}^{(l,~t)}_q}{C_{q,q}};\\
\frac{u}{2KM{\beta}^{(l,~t)}_q}-\frac{1}{C_{q,q}}, &~\text{otherwise},\end{cases}      
\end{align}  
where $u$ is the Lagrangian multiplier of constraint \eqref{eq_optimization_P6_2}, obtained by bi-section search.

The entire algorithm to solve problem \eqref{eq_optimization_P3} is summarized in Algorithm \ref{BCD for Gaussian input}. 
\begin{algorithm}[t] 
 $\textbf{Input}$ $l \leftarrow 0, ~ \overline{\mathbf{p}}^{(0)}, ~ {\boldsymbol{\sigma}}^{(0)}_R,~ {\epsilon}_S>0$;\\
 $\textbf{Output}$ $\overline{\mathbf{p}}, ~{\boldsymbol{\sigma}}$; \\
 $\textbf{Repeat}$ \\
 \begin{enumerate}
   \item Compute $\overline{\mathbf{p}}^{(l+1)}$ in problem \eqref{eq_optimization_P4} using SCP
      \begin{enumerate}
            \item[]  Initialize $t\leftarrow 0$; $\overline{\mathbf{p}}^{(l+1,~0)}\leftarrow \overline{\mathbf{p}}^{(l)}$
            \item[] \textbf{while $t\leq 10^2$}
             \begin{enumerate}
                 \item[]  Compute $\mathbf{B}^{(l+1,~t)} $  and $\boldsymbol{\alpha}^{(l+1,~t)} $ in \eqref{eq_matrixB1}-\eqref{eq_vectoralpha1}

             \item[]  Compute $\overline{\mathbf{p}}^{(l+1,~t+1)}$ using \eqref{eq_optimization_P5}

             \item[]  $t\leftarrow t+1$; $\overline{\mathbf{p}}^{(l+1)} \leftarrow \overline{\mathbf{p}}^{(l+1,~t+1)}$
                  \end{enumerate}
           \item[] Iteration complexity $\mathcal{O}\left(K^3M^3|\kappa| \right)$ 
         \end{enumerate}
   \item Compute ${\boldsymbol{\sigma}}^{(l+1)}$ in problem \eqref{eq_optimization_P6} using SCP
    \begin{enumerate}
            \item[] Initialize $t\leftarrow 0$; ${\boldsymbol{\sigma}}^{(l+1,~0)}\leftarrow {\boldsymbol{\sigma}}^{(l)}$
            \item[] \textbf{while $t\leq 10^2$}
            \begin{enumerate}
            \item[]  Compute $\boldsymbol{\beta}^{(l+1,~t)} $ in \eqref{eq_optimization_P62} and ${\boldsymbol{\sigma}}^{(l+1,~t+1)}$ in \eqref{eq_OP_CF_62}

             \item[]  $t\leftarrow t+1$; ${\boldsymbol{\sigma}}^{(l+1)} \leftarrow {\boldsymbol{\sigma}}^{(l+1,~t+1)}$
                  \end{enumerate}
               \item[] Iteration complexity    $\mathcal{O}\left( KM\right)$
         \end{enumerate}

   \item $l\leftarrow l+1$; $\overline{\mathbf{p}}\leftarrow \overline{\mathbf{p}}^{(l+1)}$; ${\boldsymbol{\sigma}}\leftarrow{\boldsymbol{\sigma}}^{(l+1)}$

   \item Quit if $|\overline{\nbbP}^{(l+1)}_{\rm UB}\left(\overline{\mathbf{p}},~\boldsymbol{\sigma}\right) -\overline{\nbbP}^{(l)}_{\rm UB}\left(\overline{\mathbf{p}},~\boldsymbol{\sigma}\right)|< {\epsilon}_S$
 \end{enumerate}
\caption{BCD for Gaussian input in \eqref{eq_optimization_P3}}
\label{BCD for Gaussian input}
\end{algorithm}

\subsection{Decoupling optimization}
\label{sec_Gaussian_decouple}
Algorithm \ref{BCD for Gaussian input} can be time-consuming given large $K$ and $M$ because of the large-dimensional variables and coefficients. However, inspired by the optimal power allocation in section \ref{sec_BPSK}, we give a simplified method by decoupling the optimization of Gaussian input distribution across time domain (OFDM symbols) and frequency domain (OFDM subcarriers), namely the decoupling method. Specifically, we assume that the distribution of CSs on each OFDM symbol is a scaled version of a common distribution $\overline{\mathbf{p}}_{\mathrm{K}}\in\mathbb{R}^{2K}$ and $\boldsymbol{\sigma}_{\mathrm{K}}\in\mathbb{R}^{2K}$ (accounting for the real and imaginary part).  Mathematically, the final distribution of $X[k,~m]$  {in \eqref{eq_X_ID} }has the form:
\begin{align}
\label{eq_decoupl_expression}
\overline{\mathbf{p}}=\mathbf{F}_{\mathrm{M}}\overline{\mathbf{p}}_{\mathrm{K}}, ~
{\boldsymbol{\sigma}}=\mathbf{F}_{\mathrm{M}}\boldsymbol{\sigma}_{\mathrm{K}},
\end{align}
where $\mathbf{F}_{\mathrm{M}}=\begin{bmatrix}
 \mathbf{p}\otimes\mathbf{I}_{K}&\mathbf{0}_{KM\times K}\\
\mathbf{0}_{KM\times K}& \mathbf{p} \otimes\mathbf{I}_{K}
\end{bmatrix} \in\mathbb{R}^{2KM\times 2K}$ with $\mathbf{p}\in \mathbb{R}^M$ being the optimal power allocation across OFDM symbols  in Section \ref{sec_BPSK}.  {Correspondingly, for the power allocation to be optimized across OFDM subcarriers, we have ${\mathbf{1}}^T\mathbf{\overline{p}}_{\mathrm{K}}\leq 1$ as a substitute of the power constraint and adopt the algorithm in Section \ref{sec_Gaussian_opt}.} 

\textcolor{blue}{In this context, the decoupling method is divided into two steps, (1) obtaining the optimal OFDM symbol power allocation, $\mathbf{p}$, directly from Algorithm \ref{ADMM}, and (2) obtaining the optimal input distribution across OFDM sub-carriers, $\mathbf{p}_\mathrm{K}$ and $\boldsymbol{\sigma}_\mathrm{K}$, through Algorithm \ref{BCD for Gaussian input} with slight modifications.} Consequently, compared with optimizing $\overline{\mathbf{p}}$ and $\boldsymbol{\sigma}$ of totally $4KM$ variables in section \ref{sec_Gaussian_opt}, after decoupling, we only need to optimize ${\mathbf{p}}$, $\overline{\mathbf{p}}_{\mathrm{K}}$ and $\boldsymbol{\sigma}_{\mathrm{K}}$ which are totally $M+4K$ variables.

\textcolor{blue}{Next, we provide details about how to modify Algorithm \ref{BCD for Gaussian input} for the decoupling method. Firstly, we re-express the objective function  $\overline{\nbbP}_{\rm UB}\left(\overline{\mathbf{p}},~\boldsymbol{\sigma}\right)$ in  \eqref{eq_averaged_outlier3} as a function of  $\overline{\mathbf{p}}_{\mathrm{K}}$ and $\boldsymbol{\sigma}_{\mathrm{K}}$ in the following}
\begin{subequations}
\begin{align}\nonumber
\overline{\nbbP}_{\rm UB}\left(\overline{\mathbf{p}}_{\mathrm{K}},~\boldsymbol{\sigma}_{\mathrm{K}}\right)
=&\sum_{{\epsilon}\in\kappa}\left[-\sqrt{{\overline{\mathbf{p}}_{\mathrm{K}}}^T{\mathbf{F}_{\mathrm{M}}}^T\mathbf{A}_{{\epsilon}}\mathbf{F}_{\mathrm{M}}\overline{\mathbf{p}}_{\mathrm{K}}}\right.\\\label{eq_averaged_outlier_decoupled1}
&\left.+\sum_{(n,~v)\neq (0,~0)}g_{n,v,{\epsilon}}\left(\mathbf{F}_{\mathrm{M}}\overline{\mathbf{p}}_{\mathrm{K}},~\mathbf{F}_{\mathrm{M}}\boldsymbol{\sigma}_{\mathrm{K}}\right)\right]\\\label{eq_averaged_outlier_decoupled2}
=&{A}_{\mathrm{D}}^0\mathbf{1}^T\overline{\mathbf{p}}_{\mathrm{K}}+\sum_{(n,~v)\neq (0,~0),~{\epsilon}}g^0_{n,v,{\epsilon}}\left(\overline{\mathbf{p}}_{\mathrm{K}},~\boldsymbol{\sigma}_{\mathrm{K}}\right),
\end{align}
\begin{align}\mbox{with}~
{A}_{\mathrm{D}}^0=&\sum_{{\epsilon}\in\kappa}{\sqrt{ {\left( \mathbf{p} ^T\mathfrak{R}\left\lbrace\mathbf{f}^{\mathrm{D}}_{0,~{\epsilon}}\right\rbrace \right)^2}+{\left( \mathbf{p} ^T\mathfrak{I}\left\lbrace\mathbf{f}^{\mathrm{D}}_{0,~{\epsilon}}\right\rbrace \right)^2} }},\\\nonumber
g^0_{n,v,{\epsilon}}&\left(\overline{\mathbf{p}}_{\mathrm{K}},~\boldsymbol{\sigma}_{\mathrm{K}}\right)\\\label{eq_gd_nv}
=&\sqrt{{\mathbf{\overline{p}}_K}^T{\mathbf{A}}^0_{n,v,{\epsilon}}\mathbf{\overline{p}}_K-2\|\mathbf{p} \|^2\|\overline{\mathbf{p}}_{\mathrm{K}}-\boldsymbol{\sigma}_{\mathrm{K}}\|^2},\\
\nonumber\label{eq_gd_nv2}
{\mathbf{A}}^0_{n,v,{\epsilon}}=&\gamma_{v,{\epsilon}}\left[{\mathfrak{R}\left\lbrace\mathbf{f}_{n}^{\mathrm{R_0}}\right\rbrace}{\mathfrak{R}\left\lbrace\mathbf{f}_{n}^{\mathrm{R_0}}\right\rbrace}^T+{\mathfrak{I}\left\lbrace\mathbf{f}_{n}^{\mathrm{R_0}}\right\rbrace}{\mathfrak{I}\left\lbrace\mathbf{f}_{n}^{\mathrm{R_0}}\right\rbrace}^T \right]\\
&+2\|\mathbf{p} \|^2,\\
\gamma_{v,{\epsilon}}= &|\mathbf{p} ^T\mathbf{f}^{\mathrm{D}}_{v,~{\epsilon}}|^2,~
\mathbf{f}_{n}^{\mathrm{R_0}}=\mathbf{1}_2\otimes \mathbf{f}_{n}^{\mathrm{R}},
\end{align}
\end{subequations}
where  {\eqref{eq_averaged_outlier_decoupled2} comes from Appendix \ref{sec_App_Simplified}.}

Given \eqref{eq_averaged_outlier_decoupled2}, the optimal $\overline{\mathbf{p}}_{\mathrm{K}}$ and $\boldsymbol{\sigma}_{\mathrm{K}}$ using the optimal ${\mathbf{p}} $ from Algorithm \ref{ADMM} are obtained  by
\begin{mini!}
 {\overline{\mathbf{p}}_{\mathrm{K}},~\boldsymbol{\sigma}_{\mathrm{K}}}{\overline{\nbbP}_{\rm UB}\left(\overline{\mathbf{p}}_{\mathrm{K}},~\boldsymbol{\sigma}_{\mathrm{K}}\right) \text{ in }\eqref{eq_averaged_outlier_decoupled2}
,}{\label{eq_optimization_P7}}{\label{eq_optimization_P7_1}}
\addConstraint{{\mathbf{1}}^{T}\overline{\mathbf{p}}_{\mathrm{K}}\leq  1}\label{eq_optimization_P7_2new}
       \addConstraint{ \frac{1}{2KM}\sum_{m}\log~ \det \left(\mathbf{I}_{2K}+p_{m}\mathbf{C}'\mathrm{diag}\{\boldsymbol{\sigma}_{\mathrm{K}}\}\right) \geq R^{\mathrm{c}}}\label{eq_optimization_P7_12}
      \addConstraint{\overline{\mathbf{p}}_{\mathrm{K}}\succeq \boldsymbol{\sigma}_{\mathrm{K}}}
      \addConstraint{\boldsymbol{\sigma}_{\mathrm{K}} \succeq \mathbf{0}},
\end{mini!}
where $\mathbf{C}'=2K\mathrm{diag}\left\lbrace{\mathbf{1}_{2}\otimes\mathbf{h}/({B\sigma^2})}\right\rbrace\in \mathbb{R}^{2K\times 2K}$.  {\eqref{eq_optimization_P7} can be solved in the same way as that in \eqref{eq_optimization_P3} in Section \ref{sec_Gaussian_opt} where $\overline{\mathbf{p}}_{\mathrm{K}}$ and $\boldsymbol{\sigma}_{\mathrm{K}}$ are updated iteratively following a BCD structure, with only slight modifications on the coefficients to couple with the optimal $\mathbf{p} $. The details are:}

\subsubsection{Updating $\mathbf{p}_{\mathrm{K}}$}
Then, we follow the same SCP procedure as \eqref{eq_optimization_P4} in Section \ref{sec_opt_p} to update $\mathbf{p}_K$, with the only difference being the coefficients in \eqref{eq_matrixB1} and \eqref{eq_vectoralpha1} updated by  {(similarly to \eqref{eq_SCP_p})}
\begin{subequations}\label{eq_update_decouple}
\begin{align}
\nonumber\label{eq_matrixB2}\mathbf{B}_0^{(l,~t)}=&2{\|\mathbf{p} \|^2}\|1\oslash\mathbf{g}_0^{(l,~t)}\|^2\mathbf{I}_{2K}+
\big(\mathbf{F}_\mathrm{R_0}\oslash\mathbf{g}_0^{(l,~t)}\big)\big(\mathbf{F}_\mathrm{R_0}\oslash\mathbf{g}_0^{(l,~t)}\big)^T\\&+\big(\mathbf{F}_\mathrm{I_0}\oslash\mathbf{g}_0^{(l,~t)}\big)\big(\mathbf{F}_\mathrm{I_0}\oslash\mathbf{g}_0^{(l,~t)}\big)^T,\\\label{eq_aplha_decouple}
\boldsymbol{\alpha}_0^{(l,t)}=&{A}_D^0\mathbf{1}_{2K}-4\|\mathbf{p} \|^2\|1\oslash\mathbf{g}_0^{(l,~t)}\|^2\big(\overline{\mathbf{p}}_{\mathrm{K}}^{(t)}-\boldsymbol{\sigma}_{\mathrm{K}}^{(t)}\big),
\end{align}
\mbox{with}
\begin{align}
 \label{eq_matrixB22}
\mathbf{g}_0^{(l,~t)}= &\left[4{{\mathbf{q}_0^{(t)}}+8\|\mathbf{p} \|^2\|{\overline{\mathbf{p}}_{\mathrm{K}}^{(t)}}\|^2-8\|\mathbf{p} \|^2\|{\overline{\mathbf{p}}_{\mathrm{K}}^{(t)}}-\boldsymbol{\sigma}_{\mathrm{K}}^{(l)}\|^2}\right]^{1/4},\\\label{eq_matrixB23}
{\mathbf{q}_0^{(t)}}=&{\mathbf{\overline{p}}_\mathrm{K}^{(t)}}^T\big[ {\mathbf{F}_\mathrm{R_0}}\odot\big({\mathbf{\overline{p}}_\mathrm{K}^{(t)}}^T{\mathbf{F}_\mathrm{R_0}}\big)^*+ {\mathbf{F}_\mathrm{I_0}}\odot\big({\mathbf{\overline{p}}_\mathrm{K}^{(t)}}^T{\mathbf{F}_\mathrm{I_0}}\big)^*\big],\\\nonumber
 \mathbf{F}_{\mathrm{R_0/I_0}}=& \sqrt{\boldsymbol{\gamma}}\otimes\mathfrak{R/I}\left\lbrace\begin{bmatrix}
\mathbf{f}^{\mathrm{R_0/I_0}}_{0} , ~\cdots, ~\mathbf{f}^{\mathrm{R_0/I_0}}_{K_G-1}\end{bmatrix}\right\rbrace,\\
& \text{removing } \left[\gamma_{0,{\epsilon}_1},~\cdots,~\gamma_{0,{\epsilon}_{|\kappa|}}\right]\otimes\mathbf{f}^{\mathrm{R_0/I_0}}_{0},\\
 \boldsymbol{\gamma}\triangleq& \big[\gamma_{-M/2,{\epsilon}_1}, ~\cdots, ~\gamma_{-M/2,{\epsilon}_{|\kappa|}}, ~\cdots, ~\gamma_{M/2-1,{\epsilon}_{|\kappa|}}\big].
\end{align}
\end{subequations}

\subsubsection{Updating $\boldsymbol{\sigma}_{\mathrm{K}}$}
 {Each entry of $\boldsymbol{\sigma}_{\mathrm{K}}$ in \eqref{eq_optimization_P7_12} is involved in $M$ $\log\det(\cdot)$ terms with different coefficients $p_m$, which requires solving an $M$-order polynomial function if following the same KKT procedure as when handling problem \eqref{eq_optimization_P6}. However, the closed form solution in \eqref{eq_OP_CF_62} for problem \eqref{eq_optimization_P6} comes from solving a linear function after KKT (only $1$ $\log\det(\cdot)$ term is involved as in \eqref{eq_optimization_P6_2}). As a solution, we lower bound the constraint in \eqref{eq_optimization_P7_12} by a function in the same form as \eqref{eq_optimization_P6_2}, and update the lower bound to approach the real constraint through iterations.} Specifically, we have
\begin{subequations}
\begin{align}
\nonumber\label{eq_op7_LB}
\frac{1}{2KM}&\sum_{m}\log~ \det \left(\mathbf{I}_{2K}+p_{m}\mathbf{C}'\mathrm{diag}\{\boldsymbol{\sigma}_{\mathrm{K}}\}\right)\\
\geq &{{\mathbf{f}_1^{(t)}}^{T}} \log~  \left(\mathbf{I}_{2K}+\mathbf{G}\boldsymbol{\sigma}_{\mathrm{K}} \right)+f_2^{(t)},\\
\mbox{with}~
\mathbf{f}_1^{(t)}=&\frac{1}{2KM}\sum_m \frac{{p}_{m}\mathbf{C}'\big(\mathbf{I}_{2K}+\mathbf{G}\mathrm{diag}\big\lbrace\boldsymbol{\sigma}_\mathrm{K}^{(t)}\big\rbrace\big)}{\big(\mathbf{G}+{p}_{
m}\mathbf{C}'\mathbf{G}\mathrm{diag}\big\lbrace\boldsymbol{\sigma}_\mathrm{K}^{(t)}\big\rbrace\big)},\\\nonumber
f_2^{(t)}=&\frac{1}{2KM}\sum_{m}\log~\det\big(\mathbf{1}_{2K}+{p}_{m}\mathbf{C}'\mathrm{diag}\big\lbrace\boldsymbol{\sigma}_{\mathrm{K}}^{(t)}\big\rbrace\big)-\\
&{\mathbf{f}_1^{(t)}}^T\log\big(\mathbf{1}_{2K}+\mathbf{G} \boldsymbol{\sigma}_{\mathrm{K}}^{(t)}\big),\\
 \mathbf{G}=& \max\{\mathbf{p} \}\mathbf{C}'.
\end{align}
\end{subequations}
\textcolor{blue}{Details about the inequality in \eqref{eq_op7_LB} is shown in Appendix \ref{Appen_ineq}. Employing the inequality in \eqref{eq_op7_LB} gives \eqref{eq_optimization_P7} the same format as in \eqref{eq_optimization_P6} in terms of the communication rate constraint.  As a result, the KKT solution for updating $\boldsymbol{\sigma}_{\mathrm{K}}$ in \eqref{eq_optimization_P6} can be adopted similarly here, whose closed-form result is given by}
\begin{align} \color{blue}
 \label{eq_OP_CF_62} 
{\sigma}_{q}=\begin{cases}{\overline{p}_{\mathrm{K},q}}^{(l)}, & ~ u_2>\frac{{\beta}^{(l,~t)}_q\left(1+G_{q,q}\overline{p}^{(l)}_{\mathrm{K},q}\right)}{f_{1,q}^{(t)}G_{q,q}};\\
0, &~ 0< u_2 <\frac{{\beta}^{(l,~t)}_q}{f_{1,q}^{(t)}G_{q,q}};\\
\frac{u_2f_{1,q}^{(t)}}{{\beta}^{(l,~t)}_q}-\frac{1}{G_{q,q}}, &~\text{otherwise},\end{cases}      
\end{align}  
\textcolor{blue}{where $u_2$ is the Lagrangian multiplier of constraint \eqref{eq_optimization_P7_12}, obtained by bi-section search.  }

 The entire decoupling method is summarized in algorithm \ref{Decouple}.
\begin{algorithm}[t]
$\textbf{Input}$ $l\leftarrow 0, ~\overline{\mathbf{p}}_{\mathrm{K}}^{(0)},~\boldsymbol{\sigma}_{\mathrm{K}}^{(0)},~{\epsilon}_S>0$;\\
 $\textbf{Output}$ $\overline{\mathbf{p}},~ \boldsymbol{\sigma}$; \\
  \begin{enumerate}
   \item Obtain ${\mathbf{p}}$ by Algorithm \ref{ADMM};  {complexity $\mathcal{O}\left( M^2\max\left\lbrace M, |\kappa|\right\rbrace\right)$}
   \item Obtain $\overline{\mathbf{p}}_{\mathrm{K}}, ~\boldsymbol{\sigma}_{\mathrm{K}}$ by Algorithm \ref{BCD for Gaussian input}, with updated coefficients  {in \eqref{eq_update_decouple} and $P^{\max}=1$; complexity $\mathcal{O}\left(K^3\right)$} 
   \item Update $\overline{\mathbf{p}}\leftarrow \mathbf{F}_{\mathrm{M}}\overline{\mathbf{p}}_{\mathrm{K}}$, $\boldsymbol{\sigma}\leftarrow \mathbf{F}_{\mathrm{M}}\boldsymbol{\sigma}_{\mathrm{K}}$
 \end{enumerate}
\caption{Decoupling Algorithm}
\label{Decouple}
\end{algorithm}

\section{Simulations}
\label{sec:simulations}
This section  {investigates the radar performance of the OFDM DFRC waveform optimized based on our proposed metrics - UBOP for PSK input distribution and aUBOP for Gaussian input distribution. For comparison, we also consider OFDM DFRC waveforms which optimize two widely used radar metrics, the CRB \cite{liyanaarachchi2021optimized,9705498} and the RMI  \cite{bica2016mutual,shi2017power}.} Simulations are performed based on the 802.11p standard (wireless access in vehicular environments) following a similar setup as in \cite{sturm2011waveform}, with a center frequency of $24$ GHz and bandwidth of $90.909$ MHz. In most simulations, we select $K=1024$, $K_G=256$ and $M=512$, which gives a range resolution of $1.67{\rm m}$ and a velocity resolution of $2.14{\rm m/s}$. We also assume a noise power density of $-208{\rm dBW/Hz}$ at both communications and radar receivers, with a communication path loss of $108$ dB and an average radar path loss of $130$ dB (target at 106.7 m) for optimization. At the target, assume an average cross section power loss of $-10{\rm dBm^2}$. For the Gaussian input, the communication channel is an NLOS model with $18$ taps from Model B in \cite{channel}. The range(velocity) of the target is  {chosen uniformly within the maximum values that can be unambiguously estimated }(213m (218m/s)).


\subsection{OFDM with PSK input distribution (constant magnitude)}
\begin{figure}
    \centering
\    \includegraphics[scale=0.45]{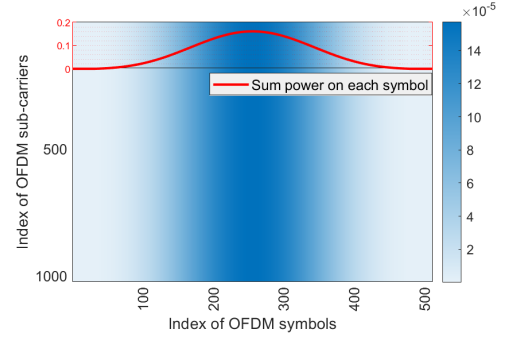}
    \caption{ {The grey-scale map 
 of the proposed optimal OFDM waveform for $K=1024$, $K_G=256$, $M=512$ at $P^{\max}=-15$ dBW (a low SNR region), with the red curve illuminating the total power (sum over subcarriers) at each OFDM symbol.}}
    \label{fig_deter_waveform}
\end{figure}

This section  {investigates the radar performance and characteristics of an OFDM DFRC waveform that minimizes UBOP with PSK input.}  {As it has been mentioned in Section \ref{sec_BPSK}, the PSK input distribution is intended to shed insight into the characteristics of an OFDM signal well-suited for on-grid delay-Doppler bin estimation, and hence the communication rate constraint is not explicitly enforced.}

First, Fig. \ref{fig_deter_waveform} sheds light on  {an OFDM DFRC waveform optimized to minimize the UBOP metric for PSK input distribution at $P^{\max}=-15$ dBW (a low SNR region). Fig. ~\ref{fig_deter_waveform} demonstrates an interesting observation that, the optimal power allocation  has uniform power allocation across subcarriers, as proved in Section \ref{sec_BPSK}. However, in the time domain, the power allocation shapes like a window (similar to Hamming or Kaiser in the low SNR), as illuminated by the red curve in Fig. ~\ref{fig_deter_waveform},  which plots the sum power on each OFDM symbol in the time domain.}  This is reasonable from the perspective of windowing to handle the problem of leakage frequencies giving fractional $v_{{\epsilon}}$. With a low receive SNR, all the sidelobes arising from the frequency leakage might lead to outlier (\textit{Remark} \ref{rem:AF} and \textit{Remark} \ref{Re:OP_AF}), in which case a window with high dynamic {delay} (such as Hamming and Kaiser) is beneficial since those dissimilar frequency components that are not aligned to the desired frequency can all be suppressed satisfyingly \cite{adams1991new}. In contrast, given a high receive SNR, the outlier mainly originates from the aligned highest sidelobe around the peak lobe, in which occasion the optimal window turns back to uniform power allocation across OFDM symbols since the rectangular window provides higher sensitivity.  {The windowing shift tendency from low SNR to high SNR is  {consistent} with  \textit{Remark} \ref{Re:OP_AF}.}

\begin{figure*}[!t]
\centering
\subfloat[]{\includegraphics[width=2.2in]{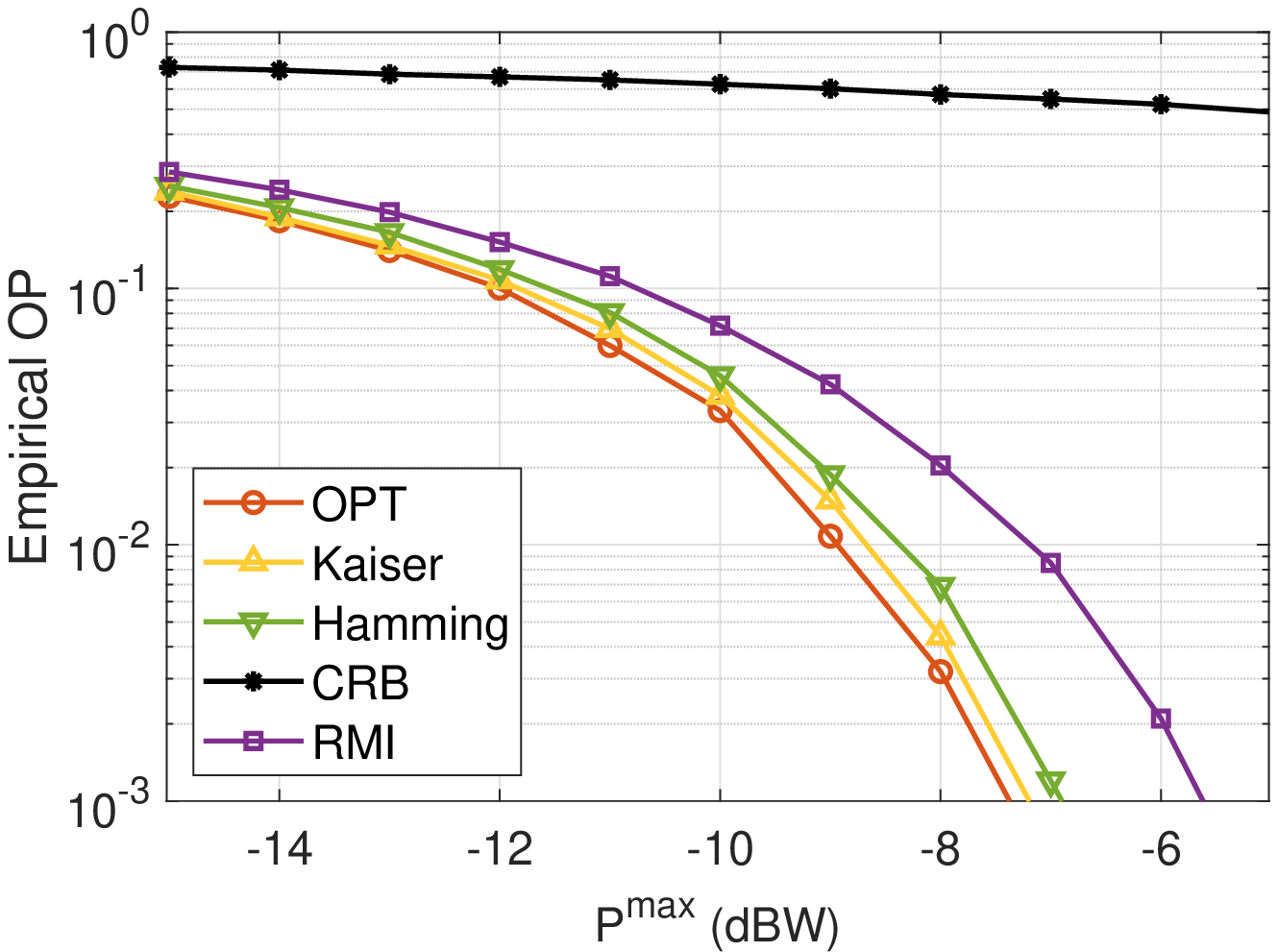}%
\label{BPSK_Po}}
\hfil
\subfloat[]{\includegraphics[width=2.2in]{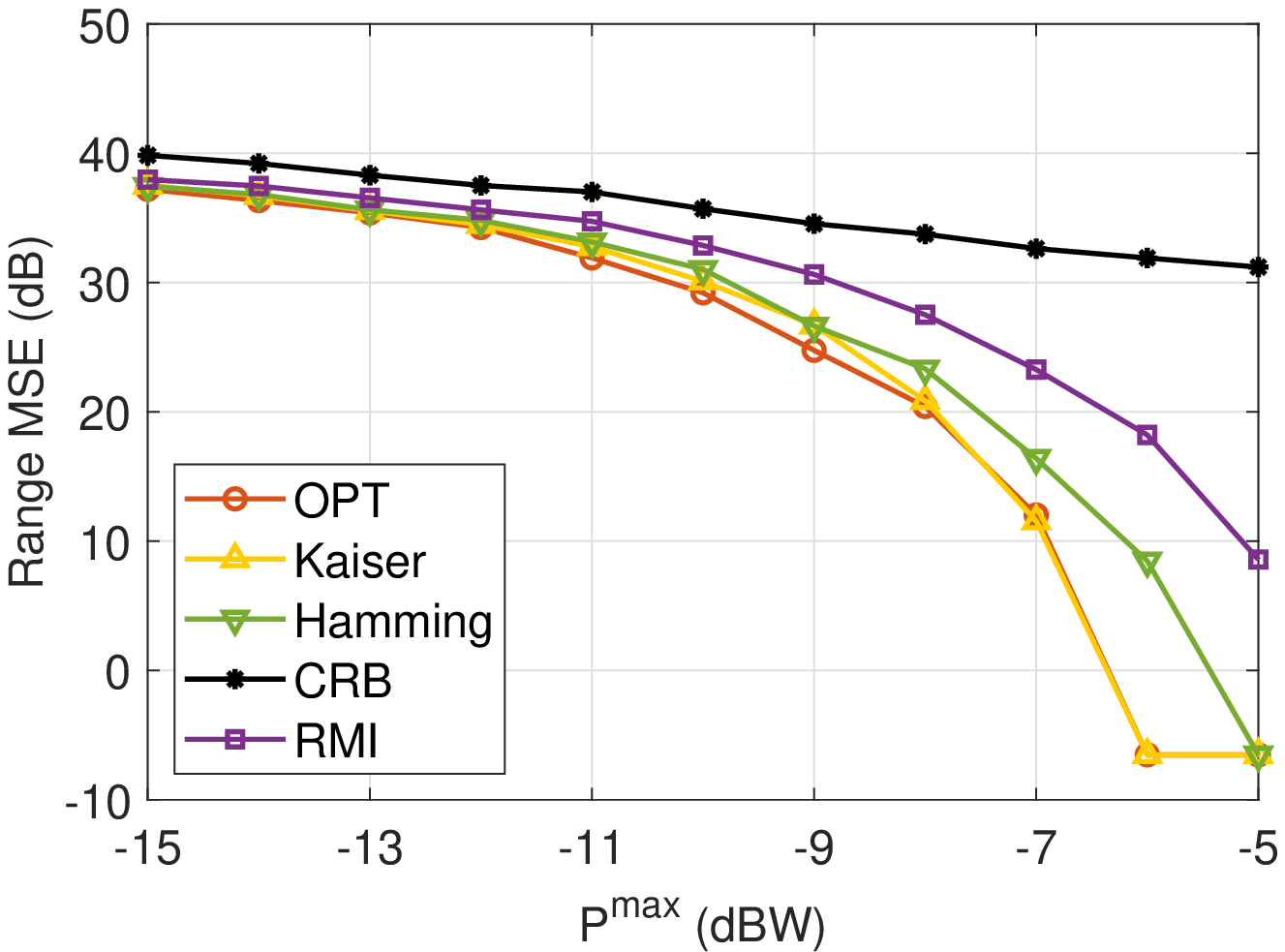}%
\label{BPSK_mse_R}}
\hfil
\subfloat[]{\includegraphics[width=2.2in]{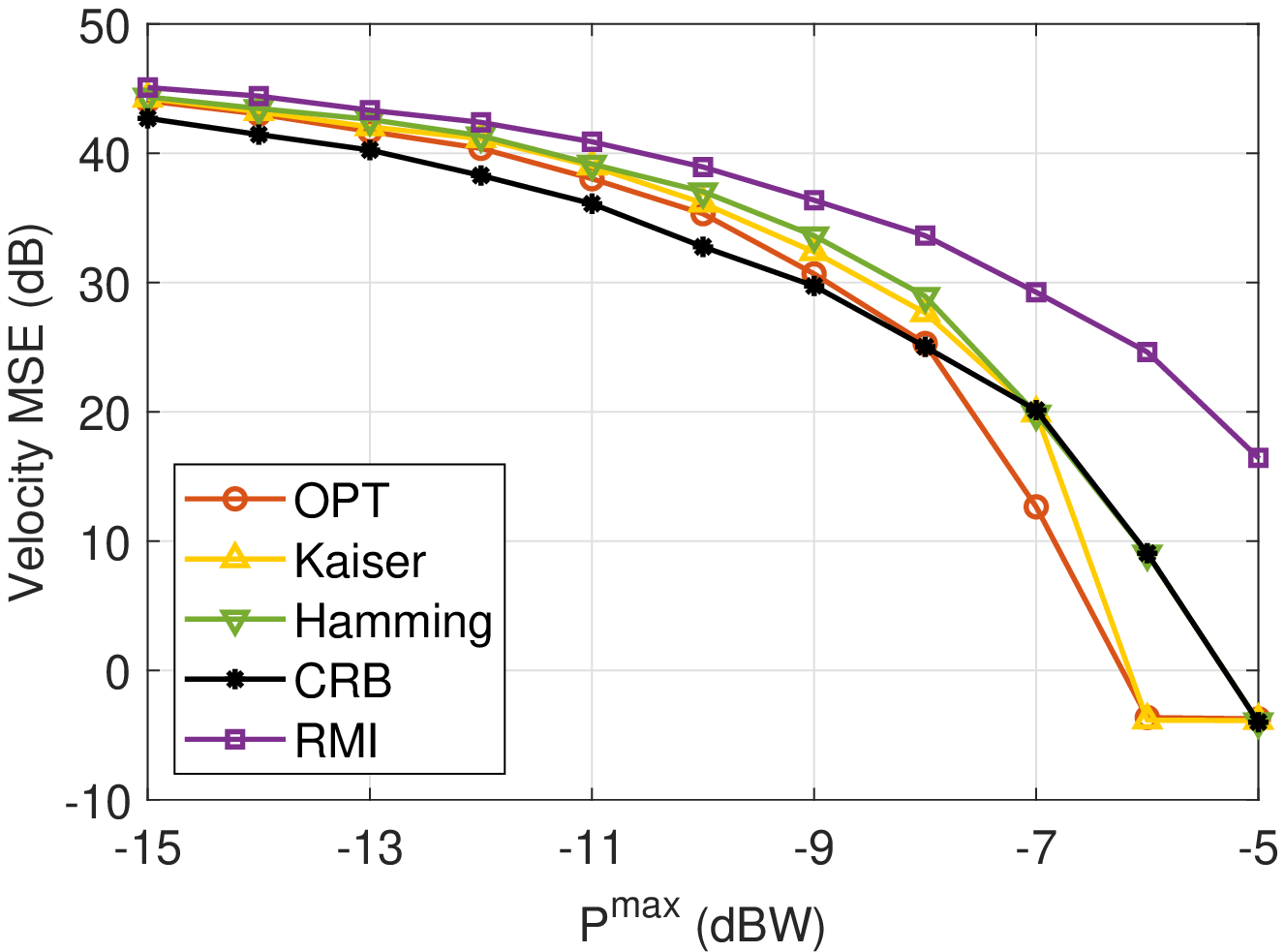}%
\label{BPSK_mse_V}}
\hfil
\caption{The empirical OP and MSE of delay-Doppler estimation between the proposed metric (OPT),  RMI, CRB and two windowed power allocation (Hamming and Kaiser) across the OFDM symbols, with BPSK input for $K=1024$, $K_G=256$, $M=512$. The worst SNR (the farthest range bin) at the radar's receiver and the SNR at the communication's receiver respectively range from $-51$ dB to $-41$ dB and from $1.4$ dB to $16.4$ dB respectively. For the subfigures, (a) OP, (b) Range MSE (representing delay), and (c) Velocity MSE (representing Doppler).}
\label{fig_deter_L32}
\end{figure*}

Given the optimal waveform intuition from Fig. \ref{fig_deter_waveform}, Fig. ~\ref{BPSK_Po} compares the  {empirical} OP between  {OFDM waveforms optimized w.r.t UBOP (red, 'OPT'), RMI (purple) and CRB (black)} as a function of the transmit power. Two windowed power allocation strategies are also introduced as an optimization-free strategy inspired by the optimal 'OPT' in Fig. \ref{fig_deter_waveform}, where uniform power allocation is adopted across subcarriers and a window (green for the 'Hamming' window and yellow for the 'Kaiser' window) is added for power allocation across symbols. Besides the OP, we also plot the MSE of the range-velocity (delay-Doppler) estimation in Fig. ~\ref{BPSK_mse_R} and Fig. ~\ref{BPSK_mse_V}, to provide more insight into the characteristics of different power allocation strategies.


Fig. ~\ref{BPSK_Po} demonstrates that our proposed power allocation 'OPT' achieves the best OP, followed by the two windowed power allocation strategies, 'Kaiser' and 'Hamming', and then 'RMI' and 'CRB'. The optimality of the 'OPT' power allocation originates from a thorough consideration of a joint delay-Doppler bin estimation, especially the frequency leakage problem as it has been analyzed. Using Hamming and Kaiser windows achieves sub-optimal performance in Fig. ~\ref{BPSK_Po} since these two windows both help suppress the sidelobes in the low SNR. Also noteworthy, between 'RMI' and 'CRB', 'RMI' shows relatively better performance on delay estimation in Fig. ~\ref{BPSK_mse_R},  {because its optimal power allocation gives uniform spectrum density (the same as the optimal power allocation across subcarriers) \footnote{ {Given a pure radar system, the RMI metric intuitively benefits the received power.}}, which is coincident with the waveform preference for delay estimation.} In contrast, 'CRB' achieves relatively better Doppler MSE (even exceeds that of the 'OPT' power allocation in low SNR) in Fig. ~\ref{BPSK_mse_V}, because the 'CRB' \footnote{ {The CRB intuitively maximizes the effective bandwidth, whose optimal power allocation naturally concentrates on the edge frequencies \cite{liyanaarachchi2021optimized}.}} power allocation obtains very reliable Doppler   estimation from the first and the last subcarriers by concentrating the power upon them. However, this results in very poor delay estimation performance on the other hand.

\subsection{OFDM with Gaussian input distribution (variable magnitude)}

\begin{figure}[!t]
\centering
{\includegraphics[width=2.8in]{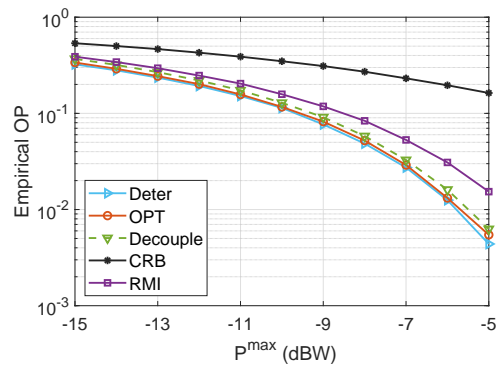}%
}
\caption{ {The empirical OP 
given $R^c=2.12$ bits/s/Hz with $K=32$, $K_G=8$, $M=16$. The red curve refers to the optimal input distribution proposed in section \ref{sec_Gaussian_opt}, shortened as 'OPT'. The green curve refers to the simplified decoupling method in section \ref{sec_Gaussian_decouple}, shortened as 'Decouple'. The cyan curve is the radar benchmark from the optimal PSK signal in Section \ref{sec_BPSK}, shortened as 'Deter'. Similarly, the purple curve refers to the 'RMI' metric and the black curve refers to the 'CRB' metric.}}
\label{fig_Gaussian_Po_K64}
\end{figure}

\begin{figure}[!t]
\centering
{\includegraphics[width=2.8in]{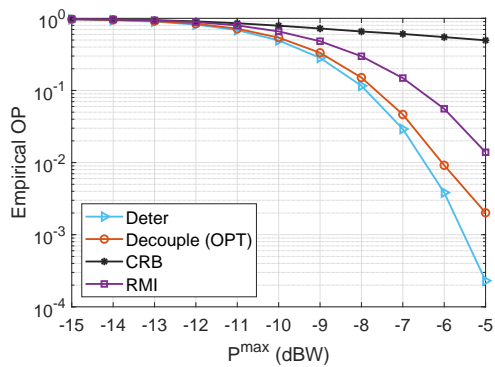}%
}
\hfil
\caption{ {The empirical OP with Gaussian input, with $K=1024$, $K_G=256$, $M=512$ (Same simulation parameters as in Fig. \ref{fig_deter_L32}.)}}
\label{fig_Gaussian_Po_K1024}\end{figure}

\begin{figure*}[!t]
\centering
\subfloat[]{\includegraphics[width=2.2in]{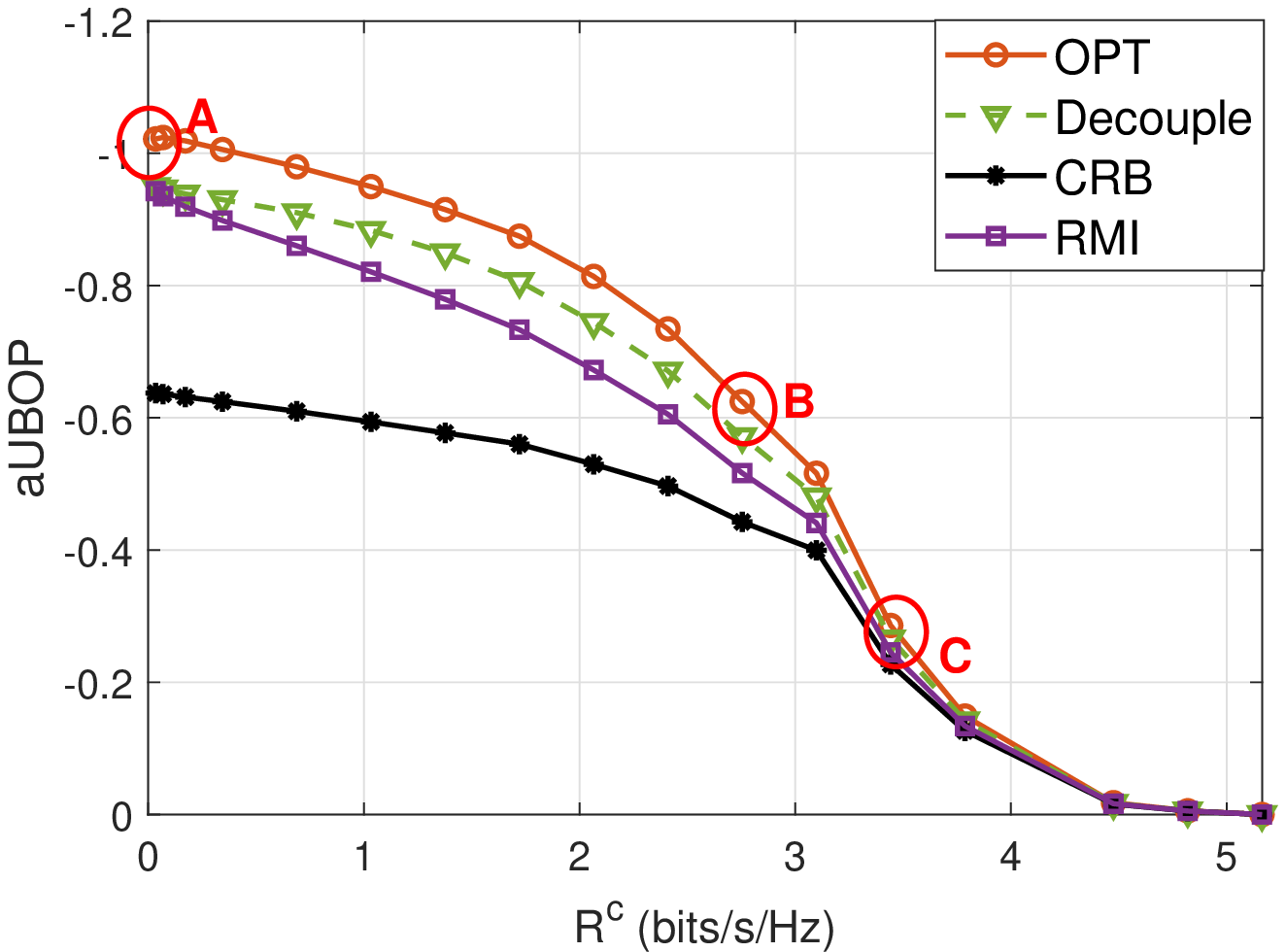}%
\label{fig_region}}
\hfil
\subfloat[]{\includegraphics[width=2.2in]{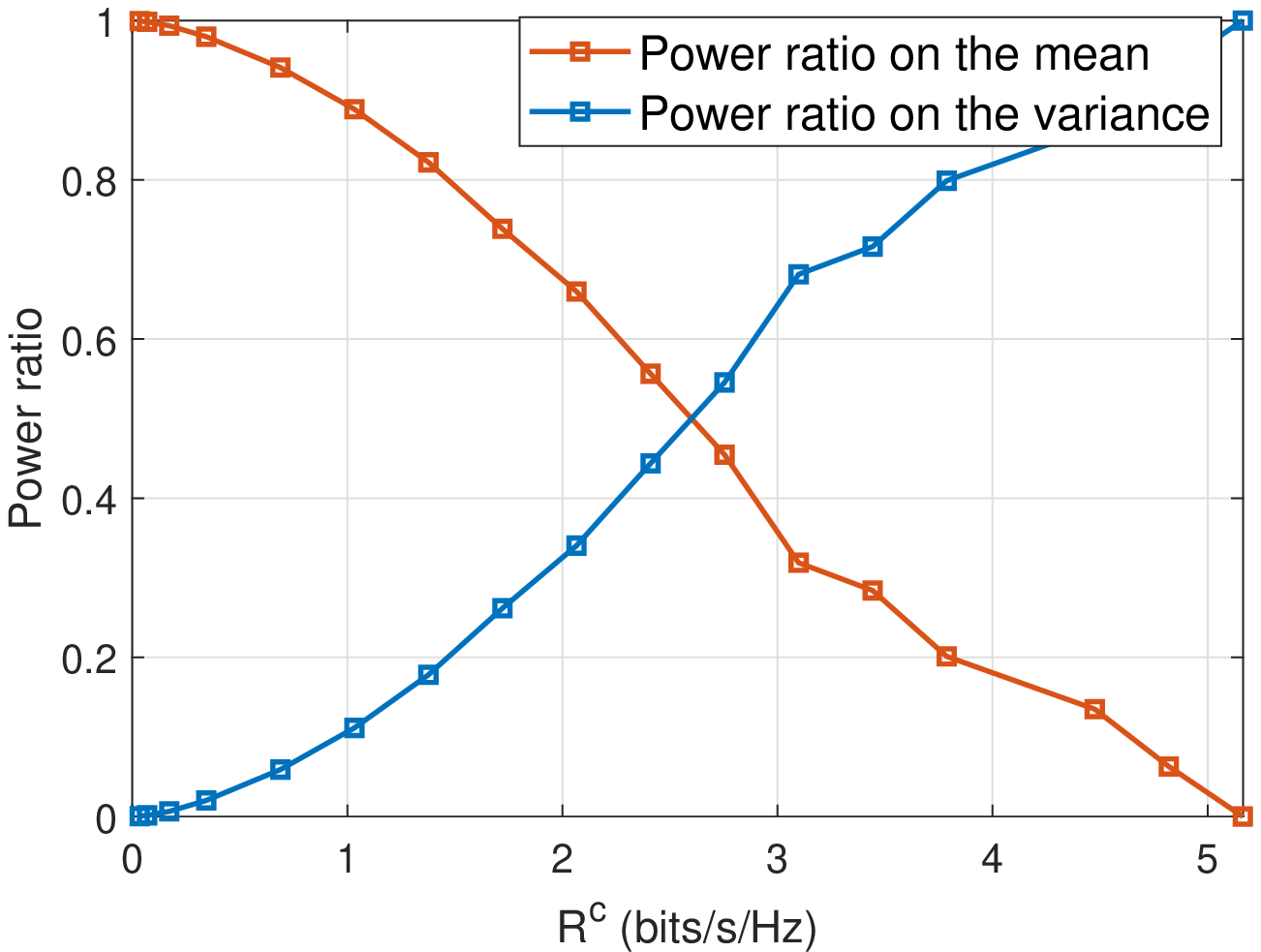}%
\label{fig_Pow_ratio}}
\hfil
\subfloat[]{\includegraphics[width=2.2in]{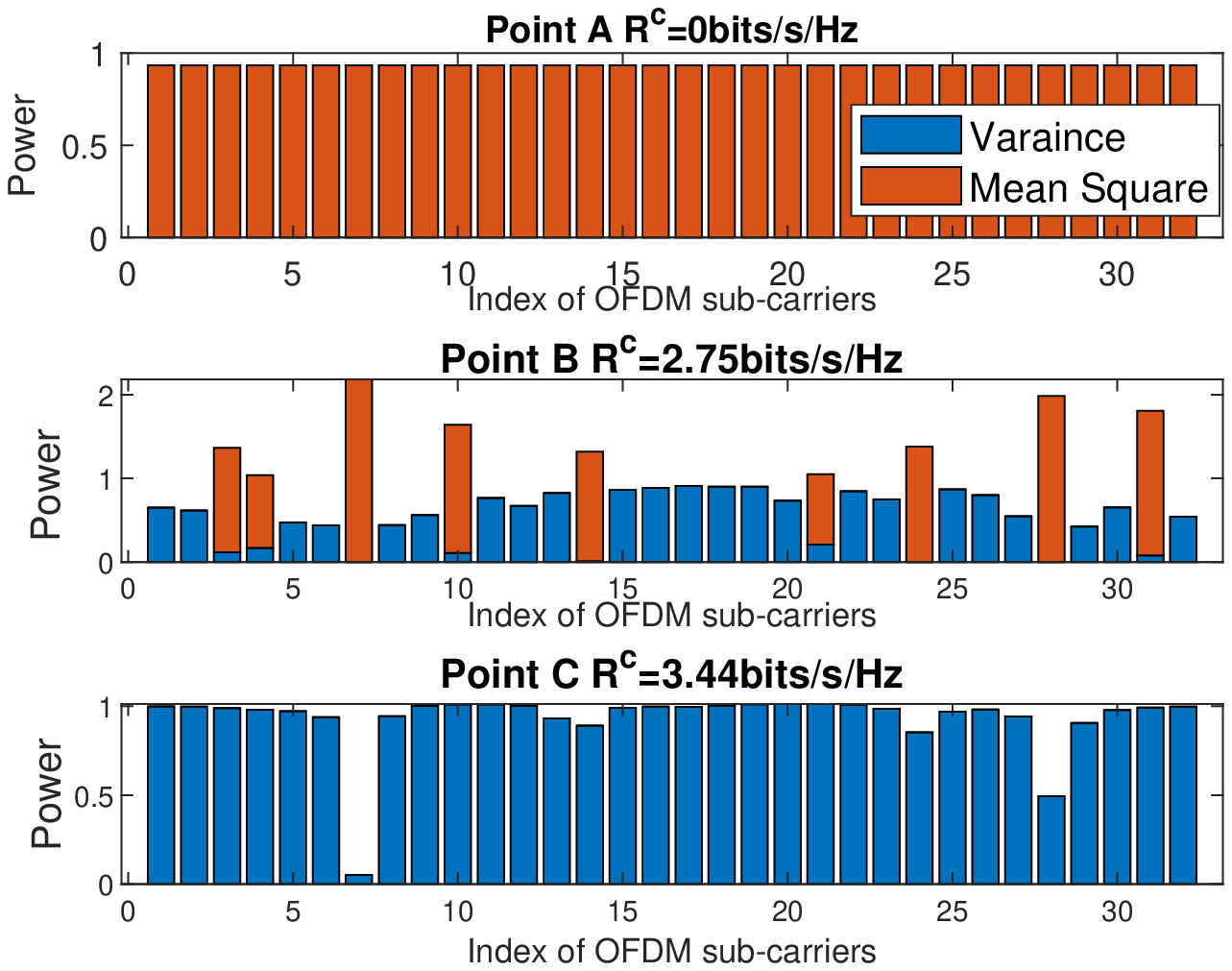}%
\label{fig_Waveform}}
\hfil
\caption{ \textcolor{blue}{The R-C trade-off in OFDM DFRC for $P^{\max}=0$ dBW, $K=32$, $K_G=8$ and $M=16$ (a) The R-C region;  (b) The power ratio of the squared symbol means (deterministic components, red) and the symbol variance (random components, blue) corresponding to the 'OPT' input distribution (red) in Fig.  \ref{fig_region}; (c) Examples of the mean power (red) and the variance power (blue) at each subcarrier (sum over OFDM symbols) in one channel realization for the 'OPT' input distribution at points 'A', 'B' and 'C' in Fig.  \ref{fig_region}. The sum of the mean power (red) and the variance power (blue) is the total power at the subcarrier. }}
\label{fig_Gaussian_R_C}
\end{figure*}

In this section, we evaluate the OP performance given Gaussian input distribution (simulated as 256-QAM). In this part, since there is no closed-form expression for the average RMI and CRB  {with communication constraints}, we substitute these two metrics with their sub-optimal solutions\footnote{ {Specifically, we first decide the minimal power that is required by the Gaussian variance to satisfy the communication constraint. The rest power of 'RMI' and 'CRB' are then allocated to their Gaussian mean for the benefit of radar, following their criterion to maximize the 'RMI' and minimize the 'CRB'.}}.  {In this section, we first verify the effectiveness of the proposed aUBOP metric by comparing its radar performance with other metrics given a fixed communication constraint (Fig.  \ref{fig_Gaussian_Po_K64} and Fig.  \ref{fig_Gaussian_Po_K1024}). Later, we focus on the R-C trade-off from the aspect of the R-C region, the randomness trade-off and the optimal waveform tendency (Fig.  \ref{fig_Gaussian_R_C}).}

Fig.  \ref{fig_Gaussian_Po_K64} simulates the radar performance with a smaller number of OFDM subcarriers and symbols to first compare the proposed two inputs, the 'OPT' (red) in Algorithm \ref{BCD for Gaussian input} and the 'Decouple' (green) in Algorithm \ref{Decouple}, together with the sub-optimal 'RMI' (purple) and 'CRB' (black) solutions. We also plot the optimal 'PSK' input as a benchmark ('Deter', cyan) to show the effect of the variable magnitudes of Gaussian CSs. Firstly, by comparing with the 'Deter' input, Fig.  \ref{fig_Gaussian_Po_K64} shows that variable magnitudes degrade the radar performance.  Then, when comparing between the four Gaussian inputs, Fig.  \ref{fig_Gaussian_Po_K64} verifies that our proposed 'OPT' input achieves the best overall performance, followed by the 'Decouple' input which has a close OP performance but with much lower complexity. To further verify the efficiency of the 'Decouple' input, we also plot in Fig.  \ref{fig_Gaussian_Po_K1024} with a normal OFDM set-up of $1024$ subcarriers and $512$ symbols, where the 'Decouple' input still outperforms the 'RMI' input and the 'CRB' input significantly. 

 {Fig.  \ref{fig_Gaussian_R_C} highlights the R-C trade-off in OFDM DFRC from different aspects (1) the R-C region; (2) the R-C trade-off w.r.t. the randomness of the 'OPT' input distribution; (3) a general trend of the 'OPT' input distribution, e.g., the power allocation strategy in the spectrum domain given different communication constraints. Specifically }

      {For (1), Fig.  \ref{fig_Gaussian_R_C}a depicts the average R-C region over $300$ channel realizations. It again shows that our proposed metrics achieve a larger R-C region than the other metrics, with the optimal being 'OPT', closely followed by the simplified 'Decouple' method. The 'CRB' is the worst since it does not fit our scenario of on-grid parameter estimation (as in Remark \ref{rem:CRB}).}

     {For (2), given the R-C region in Fig.  \ref{fig_Gaussian_R_C}a, Fig.  \ref{fig_Gaussian_R_C}b depicts the power ratio occupied by the symbol means (red) and symbol variance (blue) respectively, corresponding to the 'OPT' input distribution in Fig.  \ref{fig_Gaussian_R_C}a. It shows that, with stricter constraints on the achievable rate, the power budget is shifting from the symbol means to the symbol variance.  This illuminates the R-C trade-off from the perspective of statistics, i.e., the deterministic-random trade-off, with the former being favoured by radar and the latter being favoured by communications.}

     {For (3),  Fig.  \ref{fig_Gaussian_R_C}c sheds light on a general trend in the power allocation strategy in the spectrum domain, taking examples at points 'A' (no communication constraint), 'B' (a relatively strict communication constraint) and 'C' (the strictest achievable communication constraint in the channel realization), from the top to the bottom.    From 'A' to 'C', giving increasing achievable rate constraints, the total power is transferred from symbol means (red) to symbol variance (blue), which is coincident with the deterministic-random trade-off in (2). Specifically, 'A' emphasizes that the best radar performance is achieved by uniform power allocation across the subcarriers to the symbol mean, and 'C' is the input distribution that achieves the highest communication rate for the selected channel realization (hence whose power is all allocated to the symbol variance in a water-filling way). 'C' shows that subcarriers $17$, $19$, etc, have strong channel gains for communications and hence are allocated more power on their variances, whereas subcarriers $7$, $28$ have weak channel gains for communications on the contrary. Interestingly, if comparing 'B' with 'C' then, we observe a tendency that,  subcarriers with strong communication channel gains (e.g., $17$, $19$, etc) tend to have large variances to satisfy the achievable rate constraint in a power-efficient way, with the rest power budget allocated to the symbol means prior to the remaining subcarriers (e.g., $7$, $28$).  This roughly guarantees a relatively uniform power allocation across OFDM subcarriers, and is similar to the power allocation strategy for the constant magnitude input distribution (PSK).  }



\section{Conclusions}
\label{sec:concl}
This paper investigates the optimal input distributions of random CSs in OFDM DFRC.  {We introduce the OP as the radar metric in OFDM DFRC and show its appropriateness through analysis and simulations. The OP is then upper-bounded for optimization tractability and is a function of the magnitude of CSs, which inspires us to consider two classes of CS input distributions, (1) constant magnitude input distribution (i.e., PSK), and (2) variable magnitude input distribution (i.e., non-zero mean Gaussian). The OP is optimized under both scenarios, and low-complexity algorithms with closed-form solutions are proposed. Interestingly, through analysis and simulations, we show that for both scenarios, the optimized power allocation features a uniform power allocation across OFDM subcarriers and a windowed power allocation across OFDM symbols. For (2), the optimized input distribution exhibits a randomness trade-off, i.e., a higher achievable rate constraint for communications forces a transfer of the available power budget from the symbol means (to favour radar) to the symbol variance (to favour communications). More specifically, for subcarriers having high channel gain for communications, power is concentrated on the symbol variance in order to satisfy the communication constraints efficiently. On the other hand, for subcarriers with weak channel gains for communications, power is concentrated on the symbol means to favour radar performance. }

\appendices
\section{Derivation of the OP}
\label{sec_App_OP}

 {The probability of   $\nbbP\left[  |\gamma_{n,v,{\epsilon}}|>|\gamma_{0,0,{\epsilon}}|\right]$ in \eqref{ineq:ub} is derived based on \cite{biyari1993statistical}. Given a pair of complex Gaussian random variables $[\gamma_{0,0,{\epsilon}}, \gamma_{n,v,{\epsilon}}]$ with the mean $[r_{0,0,{\epsilon}}, r_{n,v,{\epsilon}}]$, we have} 
\begin{subequations}
\begin{align}
\label{eq_Pr_An2_AP}
&\nbbP\left[A_{n,v,{\epsilon}}\right]=\nbbP\left[|\gamma_{n,v,{\epsilon}}|>|\gamma_{0,0,{\epsilon}}|\right]=\nbbP\left[|\gamma_{0,0,{\epsilon}}|^2-|\gamma_{n,v,{\epsilon}}|^2<0\right]\\\nonumber
\approx&Q\left\lbrace\nu_1(n,~v),\nu_2(n,~v)\right\rbrace \\\label{eq_app1_1}-
&\frac{1}{2}I_0\left(\nu_1(n,~v)\nu_2(n,~v)\right)\exp\left\lbrace-\left(\nu_1(n,~v)^2+\nu_2(n,~v)^2\right)/2\right\rbrace\\\label{eq_Pr_An21_AP}
=&Q\left\lbrace\nu_1(n,~v),\nu_2(n,~v)\right\rbrace-\frac{1}{2}\exp\left\lbrace-\frac{|r_{0,0,{\epsilon}}|}{2\sigma^2}\right\rbrace I_0\left(\frac{|r_{n,v,{\epsilon}}|}{2\sigma^2}\right)\\\label{eq_Pr_An21_AP}
\approx&\frac{1}{2}\exp\big\lbrace-\frac{|r_{0,0,{\epsilon}}|}{2\sigma^2}\big\rbrace I_0\big(\frac{|r_{n,v,{\epsilon}}|}{2\sigma^2}\big),
\end{align}
\begin{align}
 \mbox{with}~\nu_1(n,~v)=&\sqrt{(r_{0,0,{\epsilon}}-R_{n,v,{\epsilon}}^{1/2})/(2\sigma^2)},\\
\nu_2(n,~v)=&\sqrt{(r_{0,0,{\epsilon}}+R_{n,v,{\epsilon}}^{1/2})/(2\sigma^2)},\\
R_{n,v,{\epsilon}}=&|r_{0,0,{\epsilon}}|^2-|r_{n,v,{\epsilon}}|^2,
\end{align}
\end{subequations}
 {where \eqref{eq_app1_1} comes from \cite{biyari1993statistical}, and \eqref{eq_Pr_An21_AP} is a further simplification adopting Marcum-Q lower bound from \cite{4753297}. } 

The Bessel function involved in \eqref{eq_Pr_An21_AP} will introduce high complexity when being used for optimization in the following. Hence, we upper bound \eqref{eq_Pr_An21_AP} as:
\begin{subequations}
\begin{align}
\label{eq_Pr_An_UB}
&\nbbP\left[A_{n,v,{\epsilon}}\right]\approx\frac{1}{2}\exp\left\lbrace-\frac{r_{0,0,{\epsilon}}}{2\sigma^2}\right\rbrace I_0\left(\frac{|r_{n,v,{\epsilon}}|}{2\sigma^2}\right)
\\
=&\frac{1}{2}\exp\left\lbrace-\frac{r_{0,0,{\epsilon}}}{2\sigma^2}\right\rbrace \sum_{p=0}^{\infty}\frac{1}{p!p!}\left(\frac{|r_{n,v,{\epsilon}}|}{4\sigma^2}\right)^{2p}\\\label{eq_Pr_An_UB3}
\leq &\frac{1}{2}\exp\big\lbrace-\frac{r_{0,0,{\epsilon}}}{2\sigma^2}\big\rbrace \big[\sum_{p=0}^{\infty}\frac{1}{p!}\left(\frac{|r_{n,v,{\epsilon}}|}{4\sigma^2}\right)^{p}\big]^2
=\frac{1}{2}\exp\big\lbrace-\frac{r_{0,0,{\epsilon}}-|r_{n,v,{\epsilon}}|}{2\sigma^2}\big\rbrace,
\end{align}
\end{subequations}
where the inequality in \eqref{eq_Pr_An_UB3} comes from  $\sum_k a_k^2\leq \left( \sum_k a_k\right)^2$ for $a_k\geq 0$  $\forall \:k$, and \eqref{eq_Pr_An_UB3} comes from the Taylor expansion of exponential functions.

\section{Proof of the Closed-form Solution for \eqref{eq_BCD_r3}}
\label{sec_APP_OPT}
For $\mathbf{r}^{(l+1,~t+1)}=\arg\min \left\lbrace{\boldsymbol{\alpha}^{(l+1,~t)}}^T\mathbf{r}^{|\cdot|}+\frac{\rho}{2}\|\mathbf{r}-\mathbf{F}^\mathrm{D}\mathbf{p} ^{(l)}\|^2 \right\rbrace
$, we have
\begin{subequations}
\begin{align}
& {\boldsymbol{\alpha}^{(l+1,~t)}}^T\mathbf{r}^{|\cdot|}+\frac{\rho}{2}\|\mathbf{r}-\boldsymbol{\beta}^{(l)}\|^2  =\sum_q \alpha^{(l+1,~t)}_q|r_{q}|+\frac{\rho}{2}|r_{q}-\beta^{(l)}_q|^2\\\nonumber\label{eq_r2_App3}
=&\sum_q \alpha^{(l+1,~t)}_v\sqrt{\mathfrak{R}\left\lbrace r_{q} \right\rbrace^2+\mathfrak{I}\left\lbrace r_{q} \right\rbrace^2}+\frac{\rho}{2}\left( \mathfrak{R}\left\lbrace r_{q} \right\rbrace^2+\mathfrak{I}\left\lbrace r_{q} \right\rbrace^2\right) \\
&-
\rho\left(\mathfrak{R}\left\lbrace r_{q} \right\rbrace\mathfrak{R}\left\lbrace \beta^{(l)}_{q} \right\rbrace+ \mathfrak{I}\left\lbrace r_{q} \right\rbrace\mathfrak{I}\left\lbrace \beta^{(l)}_{q} \right\rbrace\right)+\frac{\rho}{2}|\beta^{(l)}_q|^2\\\label{eq_r2_App4}
\geq&\sum_q \frac{\rho}{2}|r_{q}|^2+\left(\alpha^{(l)}_q-\rho|\beta^{(l)}_q|\right)|r_{q}|+\frac{\rho}{2}|\beta^{(l)}_q|^2,
\end{align}
\end{subequations}
where $\boldsymbol{\beta}^{(l)}=\mathbf{F}^\mathrm{D}\mathbf{p} ^{(l)}$. \eqref{eq_r2_App4} applies the Cauchy-Schwarz inequality and reaches the minimal if and only if $\mathfrak{R}\left\lbrace r_{q} \right\rbrace/\mathfrak{I}\left\lbrace r_{q} \right\rbrace=\mathfrak{R}\left\lbrace \beta^{(l)}_{q} \right\rbrace/\mathfrak{I}\left\lbrace \beta_{q} \right\rbrace$. Hence, the minimal value of \eqref{eq_r2_App4} is achieved by $\left[\mathbf{F}^\mathrm{D}\mathbf{p} ^{(l)}-{\angle\left(\mathbf{F}^\mathrm{D}\mathbf{p} ^{(l)}\right)\odot\boldsymbol{\alpha}^{(l)}}/{\rho}\right]^+$.

\section{Proof of \eqref{eq_averaged_outlier3}}
\label{sec_App_2nd_monents}
For \eqref{eq_averaged_outlier2}, we have  {for the first term (peak lobe related)}
 \begin{align}
\label{eq_app_confirm1}
\mathbb{E}_{X[k,~m]}\left\lbrace r_{0,0,{\epsilon}}\right\rbrace&=\sum_{k,~m}\mathbb{E}_{X[k,~m]}\left\lbrace p_{k,~m}\exp\left\lbrace {j2\pi mv_{{\epsilon}}}/{M} \right\rbrace \right\rbrace\\
&=\sum_{k,~m} \overline{p}_{k,~m}\exp\left\lbrace {j2\pi mv_{{\epsilon}}}/{M} \right\rbrace.
\end{align}
 
Then, we take expectation over $v_{{\epsilon}}$ by sampling, similarly to the solution in Section \ref{sec_BPSK}
 \begin{align}
\nonumber\label{eq_app_confirm2}
&\mathbb{E}_{ {{\epsilon}}}\left\lbrace\left\vert\mathbb{E}_{X }\left\lbrace r_{0,0,{\epsilon}}\right\rbrace \right\vert\right\rbrace=\frac{1}{|\kappa|}\sum_{v_{{\epsilon}}}|\sum_{k,~m} \overline{p}_{k,~m}\exp\left\lbrace {j2\pi mv_{{\epsilon}}}/{M} \right\rbrace|\\
&=\frac{1}{|\kappa|}\sum_{v_{{\epsilon}}}\sqrt{\overline{\mathbf{p}}^T\mathbf{A}_{{\epsilon}}\overline{\mathbf{p}}},
\end{align}
 with $\mathbf{A}_{{\epsilon}}$ in \eqref{eq_g_nv2}.

 For $(n,~v)\neq (0,~0)$ in \eqref{eq_averaged_outlier2}, we first take average over $X[k,~m]$  {(denote $f^{\mathrm{D}}_{v,{\epsilon},m}f^{\mathrm{R}}_{n,k}$ by $F^{(n,v,{\epsilon})}_{k,m}$)}:
\begin{subequations}
\begin{align}\nonumber& \mathbb{E}_{X }\left\lbrace|r_{n,v,{\epsilon}}|^2\right\rbrace
=\mathbb{E}_{X}\Big\{|\sum_{k,~m} p_{k,~m}F^{(n,v,{\epsilon})}_{k,m}|^2 \Big\}\\\nonumber
=& \sum_{k_1,~m_1,~k_2,~m_2}\mathbb{E}_{X }\left\lbrace p_{k_1,m_1}F^{(n,v,{\epsilon})}_{k_1,m_1} p_{k_2,m_2}{F^{(n,v,{\epsilon})}_{k_2,~m_2}}^* \right\rbrace\\\label{eq_confirm13}
& +\sum_{k,~m}\mathbb{E}_{X }\left\lbrace |p_{k,m}|^2 \right\rbrace\\\nonumber
=&  \sum_{k_1,~m_1,~ k_2,~m_2}\mathbb{E}_{X }\left\lbrace p_{k_1,m_1}\right\rbrace \mathbb{E}_{X }\left\lbrace p_{k_2,m_2}\right\rbrace F^{(n,v,{\epsilon})}_{k_1,m_1} {F^{(n,v,{\epsilon})}_{k_2,m_2}}^*\\
& -\sum_{k,~m}\left[\mathbb{E}_{X } \left\lbrace p_{k,m}\right\rbrace \mathbb{E}_{X }\left\lbrace p_{k,m}\right\rbrace-\mathbb{E}_{X }\left\lbrace |p_{k,m}|^2 \right\rbrace\right]\\\label{eq_confirm14}
=&  \sum_{k_1,~m_1, k_2,~m_2}\overline{p}_{k_1,m_1}\overline{p}_{k_2,m_2} F^{(n,v,{{\epsilon}})}_{k_1,m_1} {F^{(n,v,{\epsilon})}_{k_2,m_2}}^*\\&
-\sum_{k,~m}\overline{p}^2_{k,m} +\sum_{k,~m} Q_{k,m} \\\nonumber\label{eq_confirm3}
=& \left(\mathfrak{R}\left\lbrace {\mathbf{f}^{(n,v,{\epsilon})}}^T\right\rbrace\overline{\mathbf{p}}\right)^2+\left(\mathfrak{I}\left\lbrace {\mathbf{f}^{(n,v,{\epsilon})}}^T\right\rbrace\overline{\mathbf{p}}\right)^2\\& +2\sum_{k,~m}\left[{\overline{p}_{\mathrm{R},k, m}}^2+{\overline{p}_{\mathrm{I},k, m}}^2-{\mu_{\mathrm{R},k, m}}^4-{\mu_{\mathrm{I},k, m}}^4\right]\\\label{eq_confirm4}
=&\mathbf{\overline{p}}^T\mathbf{A}_{n,v,{\epsilon}}\mathbf{\overline{p}}-2\|\overline{\mathbf{p}}-\boldsymbol{\sigma}\|^2,
\end{align}
\end{subequations}
with $\mathbf{A}_{n,v,{\epsilon}}$ in \eqref{eq_g_nv3}, and where \eqref{eq_confirm14} is achieved since:
\begin{subequations}
 \begin{align}
\nonumber\label{eq_averaged_UB_define}
 Q_{k,~m}&=\mathbb{E}_{X[k,~m]}\left\lbrace | p_{k, m}|^2\right\rbrace\\\nonumber
&=3\overline{p}^2_{k, m}-2({\mu_{\mathrm{R},k, m}}^4+{\mu_{\mathrm{I},k, m}}^4)-4{\overline{p}_{\mathrm{R},k, m}}{\overline{p}_{\mathrm{I},k, m}}\\&=\overline{p}^2_{k, m}+2\left[{\overline{p}_{\mathrm{R},k, m}}^2+{\overline{p}_{\mathrm{I},k, m}}^2-{\mu_{\mathrm{R},k, m}}^4-{\mu_{\mathrm{I},k, m}}^4\right].
\end{align}
\end{subequations}

\section{Proof of \eqref{eq_averaged_outlier_decoupled2} in the Decoupling Method}
\label{sec_App_Simplified}
For the first term in \eqref{eq_averaged_outlier_decoupled1}
\begin{subequations}
\begin{align}
\nonumber&\sum_{{\epsilon}\in\kappa}\sqrt{{\overline{\mathbf{p}}_{\mathrm{K}}}^T{\mathbf{F}_{\mathrm{M}}}^T\mathbf{A}_{{\epsilon}}\mathbf{F}_{\mathrm{M}}\overline{\mathbf{p}}_{\mathrm{K}}}\\
=&\sum_{{\epsilon}\in\kappa}\footnotesize{\sqrt{{\overline{\mathbf{p}}_{\mathrm{K}}}^T{\mathbf{F}_{\mathrm{M}}}^T\left[\mathfrak{R}\left\lbrace{\mathbf{f}_{0,0,{\epsilon}}}\right\rbrace\mathfrak{R}\left\lbrace{\mathbf{f}_{0,0,{\epsilon}}}\right\rbrace^T+\mathfrak{I}\left\lbrace{\mathbf{f}_{0,0,{\epsilon}}}\right\rbrace\mathfrak{I}\left\lbrace{\mathbf{f}_{0,0,{\epsilon}}}\right\rbrace^T\right]\mathbf{F}_{\mathrm{M}}\overline{\mathbf{p}}_{\mathrm{K}}}}\\\label{AP4_1}
=&\sum_{{\epsilon}\in\kappa}{\sqrt{ {\left( \mathbf{p} ^T\mathfrak{R}\left\lbrace\mathbf{f}^{\mathrm{D}}_{0,~{\epsilon}}\right\rbrace \right)^2}+{\left( \mathbf{p} ^T\mathfrak{I}\left\lbrace\mathbf{f}^{\mathrm{D}}_{0,~{\epsilon}}\right\rbrace \right)^2} }}\mathbf{1}^T\overline{\mathbf{p}}_{\mathrm{K}}= A_0^{\mathrm{D}}\mathbf{1}^T\overline{\mathbf{p}}_{\mathrm{K}},
\end{align}
where \eqref{AP4_1} comes from 
\begin{align}\nonumber&{\mathbf{F}_{\mathrm{M}}}^T\mathfrak{R}\left\lbrace{\mathbf{f}_{0,0,{\epsilon}}}\right\rbrace\mathfrak{R}\left\lbrace{\mathbf{f}_{0,0,{\epsilon}}}\right\rbrace^T\mathbf{F}_{\mathrm{M}}\\
=&\begin{bmatrix}
\left( \mathbf{p} ^T\otimes\mathbf{I}_{K}\right){\left(\mathfrak{R}\left\lbrace\mathbf{f}^{\mathrm{D}}_{0,~{\epsilon}}\right\rbrace\otimes {\mathbf{f}^{\mathrm{R}}_{0}}\right)}\\
\left( \mathbf{p} ^T\otimes\mathbf{I}_{K}\right){\left(\mathfrak{R}\left\lbrace\mathbf{f}^{\mathrm{D}}_{0,~{\epsilon}}\right\rbrace\otimes {\mathbf{f}^{\mathrm{R}}_{0}}\right)}
\end{bmatrix} \begin{bmatrix}
\left( \mathbf{p} ^T\otimes\mathbf{I}_{K}\right){\left(\mathfrak{R}\left\lbrace\mathbf{f}^{\mathrm{D}}_{0,~{\epsilon}}\right\rbrace\otimes {\mathbf{f}^{\mathrm{R}}_{0}}\right)}\\
\left( \mathbf{p} ^T\otimes\mathbf{I}_{K}\right){\left(\mathfrak{R}\left\lbrace\mathbf{f}^{\mathrm{D}}_{0,~{\epsilon}}\right\rbrace\otimes {\mathbf{f}^{\mathrm{R}}_{0}}\right)}
\end{bmatrix}^T\\
=&{\left( \mathbf{p} ^T\mathfrak{R}\left\lbrace\mathbf{f}^{\mathrm{D}}_{0,~{\epsilon}}\right\rbrace \right)^2}\mathbf{1}_{2K\times 2K}.
\end{align}
\end{subequations}

For the second term in \eqref{eq_averaged_outlier_decoupled1}
\begin{subequations}
\begin{align}\nonumber\label{eq_gd_nv3_APP}
&{\mathbf{A}}^0_{n,v,{\epsilon}}
={\mathbf{F}_{\mathrm{M}}}^T\mathbf{A}_{n,v,{\epsilon}}\mathbf{F}_{\mathrm{M}}\\
=&\mathbf{F}_{\mathrm{M}}^T\left[{\mathfrak{R}\left\lbrace\mathbf{f}_{n,v,{\epsilon}}\right\rbrace}{\mathfrak{R}\left\lbrace\mathbf{f}_{n,v,{\epsilon}}\right\rbrace}^T+{\mathfrak{I}\left\lbrace\mathbf{f}_{n,v,{\epsilon}}\right\rbrace}{\mathfrak{I}\left\lbrace\mathbf{f}_{n,v,{\epsilon}}\right\rbrace}^T+2\mathbf{I}_{2KM} \right]\mathbf{F}_{\mathrm{M}}\\\label{eq_gd_nv3_APP2}
{=}&|\gamma_{v,{\epsilon}}|^2\left[{\mathfrak{R}\left\lbrace\mathbf{f}_{n}^{\mathrm{R_0}}\right\rbrace}{\mathfrak{R}\left\lbrace\mathbf{f}_{n}^{\mathrm{R_0}}\right\rbrace}^T+{\mathfrak{I}\left\lbrace\mathbf{f}_{n}^{\mathrm{R_0}}\right\rbrace}{\mathfrak{I}\left\lbrace\mathbf{f}_{n}^{\mathrm{R_0}}\right\rbrace}^T \right]+2\|\mathbf{p} \|^2,
\end{align}
\end{subequations}
where $\gamma_{v,{\epsilon}}= |\mathbf{p} ^T\mathbf{f}^{\mathrm{D}}_{v,~{\epsilon}}|^2$ and $\mathbf{f}_{n}^{\mathrm{R_0}}=\mathbf{1}_2\otimes \mathbf{f}_{n}^{\mathrm{R}}\in\mathbb{C}^{2K}$. 

\section{Proof of the inequality in \eqref{eq_op7_LB}}\label{Appen_ineq}
\textcolor{blue}{The inequality in \eqref{eq_op7_LB} is equivalent to, for $\forall~k,~m$,}
\begin{subequations}\color{blue}
\begin{align}
\label{Appen_eq_op7_LB}
\log~  \left(1+p_{m}{C}'_{k,k}{\sigma}_{\mathrm{K},k}\right)
\geq &{f}_{1,k,m}^{(t)} \log~  \left(1+{G}_{k,k}{\sigma}_{\mathrm{K},k} \right)+f_{2,k,m}^{(t)},
\end{align}
\mbox{with}~
\begin{align}
{f}_{1,k,m}^{(t)}=& \frac{{p}_{m}{C}'_{k,k}\big(1+{G}_{k,k}{\sigma}_\mathrm{K,k}^{(t)}\big)}{\big({G}_{k,k}+{p}_{
m}{C}'_{k,k}{G}_{k,k}{\sigma}_{\mathrm{K},k}^{(t)}\big)},
\\
f_{2,k,m}^{(t)}=&\log~\big(1+{p}_{m}{C}'_{k,k}{\sigma}_{\mathrm{K},k}^{(t)}\big)-{{f}_{1,k,m}^{(t)}}^T\log\big(1+{G}_{k,k} {\sigma}_{\mathrm{K},k}^{(t)}\big),\\
{G}_{k,k}=& \max\{{p}\}{C}'_{k,k},
\end{align}
\end{subequations}
\textcolor{blue}{where \eqref{Appen_eq_op7_LB} intrinsically comes from $\log(1+c_1x)\geq c_2\log(1+c_3x)+c_4$ for $c_1\geq c_3$ with $c_2=c_1(1+c_3x_0)/(c_3+c_1c_3x_0)$ and $c_4=\log(1+c_1x_0)-c_2\log(1+c_3x_0)$ ({by making the Talyor approximation of $\log(1+c_1x)$ w.r.t the term $\log(1+c_3x)$, i.e., $\log(1+c_1x)=\log\left(1+c_1/c_3\left(e^{t}-1\right)\right)|_{t=\log(1+c_3x)}$)}. Specifically, in \eqref{Appen_eq_op7_LB}, each sum entry of $\log\left( 1+{p}_{m}{C}'_{k,k} 
{\sigma}_{\mathrm{K},k}\right)$ for $\forall$ $m$ is approximated at the point $\log\left( 1+{G}_{k,k}{\sigma}_{\mathrm{K},k} \right)$, and combining all the $\log(\cdot)$ entries obtains the final expression in \eqref{eq_op7_LB}. }

\bibliographystyle{IEEEtran}
\bibliography{IEEEabrv, references}

 \end{document}